\documentclass[aps,pra,reprint,amsmath,amssymb,superscriptaddress,onecolumn,longbibliography, notitlepage, floatfix]{revtex4-1}

\usepackage[hidelinks]{hyperref}

\usepackage{mathtools}
\usepackage{braket}
\usepackage{siunitx}
\usepackage{float}

\usepackage{xcolor}

\begin{document}
%\nocite{*}
\title{An optical neural network using less than 1 photon per multiplication}

\author{Tianyu~Wang}
\email{tw329@cornell.edu}
\affiliation{School of Applied and Engineering Physics, Cornell University, Ithaca, NY 14853, USA}

\author{Shi-Yuan~Ma}
\affiliation{School of Applied and Engineering Physics, Cornell University, Ithaca, NY 14853, USA}

\author{Logan~G.~Wright} 
\affiliation{School of Applied and Engineering Physics, Cornell University, Ithaca, NY 14853, USA}
\affiliation{NTT Physics and Informatics Laboratories, NTT Research, Inc., Sunnyvale, CA 94085, USA}

\author{Tatsuhiro~Onodera}
\affiliation{School of Applied and Engineering Physics, Cornell University, Ithaca, NY 14853, USA}
\affiliation{NTT Physics and Informatics Laboratories, NTT Research, Inc., Sunnyvale, CA 94085, USA}

\author{Brian~C.~Richard}
\affiliation{School of Electrical and Computer Engineering, Cornell University, Ithaca, NY 14853, USA}

\author{Peter~L.~McMahon}
\email{pmcmahon@cornell.edu}
\affiliation{School of Applied and Engineering Physics, Cornell University, Ithaca, NY 14853, USA}

\begin{abstract}

Deep learning has rapidly become a widespread tool in both scientific and commercial endeavors \cite{lecun2015deep}. Milestones of deep learning exceeding human performance have been achieved for a growing number of tasks over the past several years, across areas as diverse as game-playing, natural-language translation, and medical-image analysis. However, continued progress is increasingly hampered by the high energy costs associated with training and running deep neural networks on electronic processors \cite{thompson2020computational}. Optical neural networks have attracted attention as an alternative physical platform for deep learning \cite{shastri2021photonics,Wetzstein2020}, as it has been theoretically predicted that they can fundamentally achieve higher energy efficiency than neural networks deployed on conventional digital computers \cite{hamerly2019large,nahmias2020photonic}. Here, we experimentally demonstrate an optical neural network achieving $99\%$ accuracy on handwritten-digit classification using $\sim$3.2 detected photons per weight multiplication and $\sim$90\% accuracy using $\sim$0.64 photons ($\sim$\SI{2.4E-19}{\joule}  of optical energy) per weight multiplication. This performance was achieved using a custom free-space optical processor that executes matrix-vector multiplications in a massively parallel fashion, with up to $\sim$0.5 million scalar (weight) multiplications performed at the same time. Using commercially available optical components and standard neural-network training methods, we demonstrated that optical neural networks can operate near the standard quantum limit with extremely low optical powers and still achieve high accuracy. Our results provide a proof-of-principle for low-optical-power operation, and with careful system design including the surrounding electronics used for data storage and control, open up a path to realizing optical processors that require only \SI{E-16}{\joule} \textit{total} energy per scalar multiplication—which is orders of magnitude more efficient than current digital processors.

\end{abstract}

\maketitle

\section{Introduction}
Much of the progress in deep learning over the past decade has been facilitated by the use of ever-larger models, with commensurately larger computation requirements and energy consumption \cite{thompson2020computational}. The widespread commercial implementation of increasingly complex deep neural networks (DNNs) has resulted in rapid growth in the total energy consumption of machine learning, and in large-scale deployments, 80-90$\%$ of the cost is for inference processing \cite{Jassy2019reinvent}. The rate at which the energy requirements for state-of-the-art DNNs is growing is unsustainable and urgently needs to be addressed \cite{thompson2020computational}. Both software and hardware advances are important for reducing energy consumption: DNN models (the software) can be made more efficient \cite{thompson2020computational}, and hardware for executing DNN models can be made more efficient, by specializing the hardware to the type of computations involved in DNN processing \cite{sze2017efficient}.

Optical processors have been proposed as deep-learning accelerators that can in principle achieve better energy efficiency and lower latency than electronic processors \cite{shastri2021photonics,Wetzstein2020,hamerly2019large,nahmias2020photonic,caulfield2010future}. For deep learning, optical processors’ main proposed role is to implement matrix-vector multiplications \cite{shen2017deep,lin2018all}, which are typically the most computationally-intensive operations in DNNs \cite{sze2017efficient} (Fig. \ref{figure1}a).

Theory and simulations have suggested that optical neural networks (ONNs) built using optical matrix-vector multipliers can exhibit extreme energy efficiency surpassing even that of irreversible digital computers operating at the fundamental thermodynamic limit \cite{hamerly2019large}. In order to achieve an energy advantage, an optical matrix-vector multiplier needs massively parallel execution of the scalar multiplications and additions that constitute a matrix-vector multiplication \cite{hamerly2019large,nahmias2020photonic}. It has been predicted that for sufficiently large vector sizes, matrix-vector multiplication can be performed with an optical energy cost of less than 1 photon per scalar multiplication, assuming the standard quantum limit for noise \cite{hamerly2019large,nahmias2020photonic}. The energy of a single photon at visible wavelengths is on the order of \SI{E-19}{\joule}; this suggests a possibility for optical processors to have an energy advantage of several orders of magnitude over electronic processors using digital multipliers, whose energy cost is currently between \SI{E-14}{} and \SI{E-13}{\joule} per scalar multiplication with equivalent precision \cite{reuthersurvey2020,horowitz20141}. 

The key to realizing an energy advantage in an optical matrix-vector multiplier is to maximize the sizes of the matrices and vectors that are to be multiplied. With large operands, there are many constituent scalar multiplication and accumulation operations that can be performed in parallel completely in the optical domain, and the costs of conversions between electronic and optical signals can be amortized \cite{nahmias2020photonic}. Optics provides several different ways to implement operations in parallel, including using wavelength multiplexing \cite{feldmann2021parallel,xu202111}, spatial multiplexing in photonic integrated circuits \cite{shen2017deep,feldmann2021parallel,tait2019silicon,stark2020opportunities,bogaerts2020programmable,wu2021programmable}, and spatial multiplexing in 3D free-space optical processors \cite{lin2018all,miscuglio2020massively,goodman1978fully,psaltis1988adaptive,dong2019optical,chang2018hybrid,matthes2019optical,bueno2018reinforcement,spall2020fully,bernstein2021freely,zhou2021large}.

To date, across all multiplexing approaches and architectures, demonstrations of analog ONNs have involved small vector-vector dot products (as a fundamental operation in implementing convolutional layers \cite{feldmann2021parallel,xu202111} and fully connected layers \cite{spall2020fully}) or matrix-vector multiplications (for realizing fully connected layers \cite{shen2017deep}): the vectors have been limited \cite{ramey2020silicon} to sizes of at most 64. This is substantially below the scale (vector sizes \textgreater$10^3$) at which sub-photon-per-multiplication energy efficiency is predicted to be feasible \cite{hamerly2019large,nahmias2020photonic}. This is the fundamental reason that the optical energy consumption in recently demonstrated optical processors is still several orders of magnitude higher than that of theoretical predictions (\SIrange[range-phrase=-]{E-14}{E-13}{} versus \SI{E-18}{\joule} per scalar multiplication) \cite{hamerly2019large,nahmias2020photonic,xu202111,spall2020fully,ramey2020silicon}. One ONN architectural approach that is promising for near-term explorations of large-scale ONN operation is to perform spatial multiplexing in 3D free space, since a 2D cross section can contain \cite{miller2019waves}, for example, \textgreater$10^6$ modes in an area of \SI{1}{\milli \metre \squared}. While the potential for large-scale operation exists based on the available parallelism in free-space spatial modes, this potential has not yet been realized.

Here, we report on our experimental implementation of a 3D free-space optical processor that can perform matrix-vector multiplications at large scale (with vector sizes of up to 0.5 million). The large scale has enabled us to demonstrate the computation of matrix-vector products and vector-vector dot products each using less than 1 photon per scalar multiplication. With this optical matrix-vector multiplier we constructed an ONN (Fig. \ref{figure1}a) that performs image classification using less than 1 photon per scalar multiplication, matching theoretical predictions for the quantum-limited optimal efficiency of an ONN \cite{hamerly2019large}.

\section{Large-scale Optical Matrix-Vector Multiplication}

\begin{figure}[h!]
    \centering
    \includegraphics[width=0.77\textwidth]{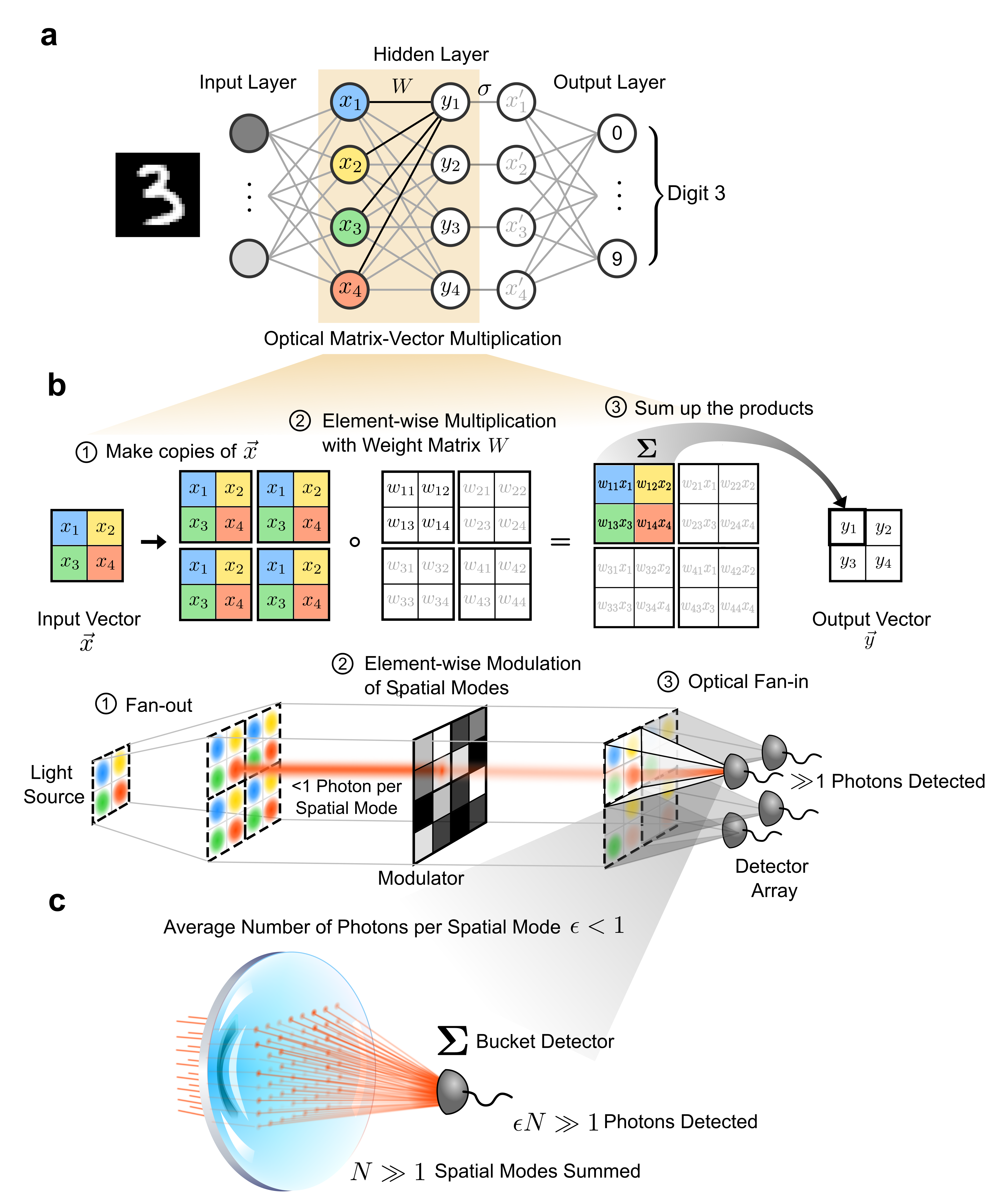}
    \caption{\textbf{A 3D free-space optical matrix-vector multiplier that uses less than one photon per scalar multiplication performed.} 
    \textbf{a,} The role of optical matrix-vector multiplication in executing the forward-pass (inference-mode) operation in a fully connected neural network. The matrix multiplications involved in propagating each layer forward are performed optically and the element-wise nonlinear activation functions are performed electronically. Each neuron in the middle (hidden) layer is color-coded to show the correspondence between the mathematically abstract neurons in (a) and their optical implementation in (b).
    \textbf{b,} A step-by-step illustration of how optical matrix-vector-multiplication can be performed by decomposing the matrix-vector operation $W \vec{x}$ into blocks of vector-vector dot products that are implemented as scalar multiplications performed in parallel, followed by summations (accumulations) performed in parallel. The depiction is for a 4$\times$4 matrix and a 4-dimensional vector, but the concept extends naturally to matrices of arbitrary dimension. The top row shows mathematically abstract operations, and the bottom row shows the corresponding physical operations with optics. “$\circ$” denotes element-wise multiplication between two matrices or vectors of the same size. Individual scalar multiplications are realized optically by encoding one operand ($x_j$) in the intensity of a single spatial mode (depicted as a beam for illustrative purposes), and the other operand ($w_{ij}$) in the transmissivity of a single pixel of a modulator: propagation of the beam through the modulator’s pixel yields the scalar multiplication, the result of which is encoded in the intensity of the transmitted light. Each summation $\sum_{j}w_{ij}x_j$ is performed optically by focusing multiple spatial modes onto a single detector. 
    \textbf{c,} An illustration of how optical matrix-vector multiplication can consume less than 1 photon per scalar multiplication when for a large vector size. A single lens is used to sum the intensities of the spatial modes encoding the element-wise products $w_{ij}x_j$ between the $i$th row of $W$ and the vector $\vec{x}$. For a vector of size $N$, there are $N$ spatial modes whose intensities are summed. If $N$ is sufficiently large then even if each individual spatial mode contains $\epsilon$ \textless 1 photon on average, the total number of photons impinging on the detector will be   $\epsilon \gg1$, allowing high signal-to-noise-ratio measurement of the summation result $y_i = \sum_{j}w_{ij}x_j$.}
    \label{figure1}
\end{figure}

The optical processor we designed and constructed uses the following scheme to perform matrix-vector multiplications $y=W\vec{x}$. Each element $x_j$ of the input vector $\vec{x}$ is encoded in the intensity of a separate spatial mode illuminated by a pixel of a light source, and each matrix element $w_{ij}$ is encoded as the transmissivity of a modulator pixel. We used an organic light-emitting diode (OLED) display as the light source and a spatial light modulator (SLM) for intensity modulation. Matrix-vector multiplications were computed in three physical steps: 1) Fan-out: The input vector’s elements were spatially arranged into a 2D block (Fig. \ref{figure1}b, top left). The 2D block representing $\vec{x}$ was replicated a number of times equal to the number of rows in the matrix $W$, and then tiled on the OLED display as shown in Fig. \ref{figure1}b (top row). 2) Element-wise Multiplication: Each OLED pixel encoding a single vector element $x_j$ was aligned and imaged to a corresponding pixel on the SLM, whose transmissivity was set to be $\propto w_{ij}$. This performs the scalar multiplication $w_{ij} x_j$ (Fig. \ref{figure1}b bottom middle). 3) Optical Fan-in: The intensity-modulated pixels from each block were physically summed by focusing the light transmitted by them onto a detector. The total number of photons impinging on the $i$th detector is proportional to the element $y_i$ of the matrix-vector product $y=W\vec{x}$ (Fig. \ref{figure1}b bottom right). We can interpret each $y_i$ as the dot product between the input vector $\vec{x}$ and the $i$th row of the matrix $W$.

This design computes all the scalar multiplications and additions involved in a matrix-vector multiplication in parallel in a single pass of light through the setup. The encoding of vector elements in optical \textit{intensity} constrains the setup to performing matrix-vector multiplications with matrices and vectors that have \textit{non-negative} elements. However, we can use the system to perform matrix-vector multiplications with matrices and vectors that have \textit{signed} elements (elements that can take both positive and negative values) by using offsets and scaling to convert the calculations to matrix-vector multiplications involving only non-negative numbers \cite{goodman1978fully} (see Supplementary Information Section 11).

Our 2D-block design can be, and was, implemented with spherical-lens systems, which are well-corrected for errors in large-field-of-view imaging (The setup schematic is shown in Supplementary Information Section 1.) The setup enabled us to align an array of 711$\times$711 pixels on the OLED display to an array of 711$\times$711 pixels on the SLM (Supplementary Information Sections 5, 6, and 7), allowing 711$\times$711=505,521 scalar multiplications to be performed in parallel. Our experimental setup was, with a single pass of light through the setup, capable of performing matrix-vector multiplications with matrices having arbitrary dimensions, subject to the total number of matrix elements being no larger than 505,521. In the special case of the matrix merely being a vector, our setup performed a single vector-vector dot product with vectors sizes up to 505,521.

The ability to perform dot products between very large vectors with our block-encoded design enabled us to achieve extremely low optical energy consumption. For each vector-vector dot product that the system computes, the summation of the element-wise products is performed by focusing the spatial modes corresponding to the element-wise products onto a single detector. If the vectors have size $N$, then $N$ spatial modes are incoherently summed on the detector. 

Consequently the detector’s output, which is proportional to the dot-product answer, has a signal-to-noise ratio (SNR) that scales as $\sqrt{N}$ under the shot-noise limit \cite{nahmias2020photonic}. If the vector size  is sufficiently large then even if the individual spatial modes each have an average photon number far less than 1, the total number of photons impinging on the detector can be much greater than 1, and so precise readout of the dot-product answer is possible (Figure \ref{figure1}c). 

\section{Precision of sub-photon dot products}

\begin{figure}[h]
    \centering
    \includegraphics[width=0.9\textwidth]{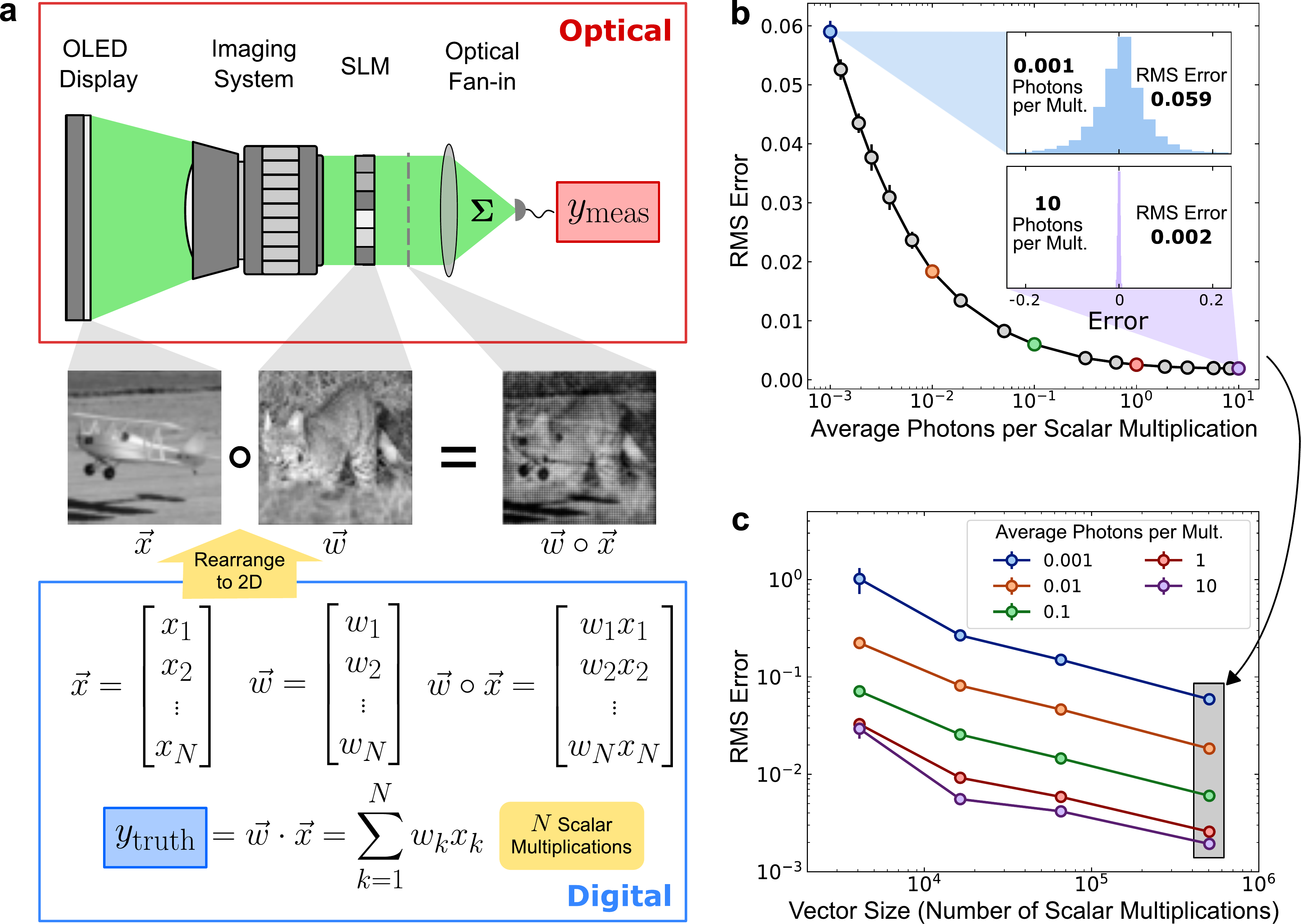}
    \caption{\textbf{Vector-vector dot products computed with high accuracy and high effective numerical precision using as few as 0.001 photons per scalar multiplication.} 
    \textbf{a,} Simplified schematic of optical setup for computation of vector-vector dot products, and characterization procedure. $N$-pixel images were used as test vectors by interpreting each image as an $N$-dimensional vector. The setup was used to compute the dot products between many different random pairs of vectors, with each computation producing a result $y_\textrm{meas}$ (top and center rows; example experimental measurement of element-wise multiplication $\vec{w} \circ \vec{x}$ was captured with a camera before optical fan-in for illustrative purposes). For each vector pair, the dot-product ground truth $y_{\textrm{truth}}$ was computed on a digital computer (bottom row). The error was calculated as $y_{\textrm{meas}} - y_{\textrm{truth}}$. \textbf{b,} The root-mean-square (RMS) error of the dot product computation as a function of the average number of detected photons per scalar multiplication. The vector length $N$ was $\sim$0.5 million (711$\times$711), which is sufficiently large that we observed an RMS error of \textless6$\%$ even when only 0.001 photons per multiplication were used, and an RMS error of \textless1$\%$ when 0.1 photons per multiplication were used. The insets show error histograms (over different vector pairs and repeated measurements) from experiments using 10 and 0.001 photons per multiplication, respectively. The error bars in the main plot show $10\times$ the standard deviation of the RMS error, calculated using repeated measurements. 
    \textbf{c,} The RMS error as a function of the vector size $N$, equal to the number of scalar multiplications needed to compute a vector-vector dot product. For each vector size, the RMS error was computed using five different photon budgets, ranging from 0.001 to 10 photons per scalar multiplication. The shaded column indicates data points that are also shown in Panel b. The error bars show $3 \times$ the standard deviation of the RMS error, calculated using repeated measurements. }
    \label{figure2}
\end{figure}

To understand how well our system performed in practice in the regime of low optical power consumption, we characterized its accuracy while varying the number of photons used. In our first characterization experiments, we computed the dot products of randomly chosen pairs of vectors (Figure \ref{figure2}a; see Methods). The results from our characterization with dot-product computations apply directly to the setting of matrix-vector multiplications with generic matrices because our setup computes matrix-vector multiplications as a set of vector-vector dot products. For a dot-product computation, the answer is a scalar, so only a single detector was used: the optical signal encoding the dot-product solution was measured by a sensitive photodetector capable of resolving single photons (see Methods), and the number of photons used for each dot product was controlled by changing the detector integration time and by inserting neutral-density filters immediately after the OLED display.

To demonstrate our setup could perform computations using less than 1 photon per scalar multiplication for large vector sizes, we measured the numerical precision of dot products between vectors each of size $\sim$0.5 million. With 0.001 photons per scalar multiplication, the error was measured to be $\sim$6$\%$ (Figure \ref{figure2}b; see Supplementary Information Section 12 for the details of RMS-error calculation); the dominant contribution to this error was from shot noise at the detector (Supplementary Information Section 8). As we increased the number of photons used, the error decreased until it reached a minimum of $\sim$0.2$\%$ at 2 photons per multiplication or higher (Figure \ref{figure2}b). We hypothesize that the dominant sources of error at high photon counts (\textgreater2 photons per multiplication) are imperfect imaging of the OLED display pixels to SLM pixels, and crosstalk between SLM pixels. We note that the experimental runs to test the performance of the system when using 0.001 photons per multiplication (which resulted in $\sim$6$\%$ error) were performed with a detector integration time of $\sim$\SI{100} {\nano \second}. This shows that matrix-vector multiplications can be performed with \textless1 photon per multiplication at a rate of at least 10 MHz, although this is merely an experimentally demonstrated \textit{lower-bound}: in principle, with sufficiently fast modulators and detectors, the system should be able to perform matrix-vector multiplications at rates \textgreater10 GHz \cite{nahmias2020photonic}. To enable comparison between the experimentally achieved analog numerical precision with the numerical precision in digital processors, we can interpret each measured analog error percentage (Figure \ref{figure2}b) as corresponding to an effective bit-precision for the computed dot product’s answer. Using the metric \textit{noise-equivalent bits} \cite{nahmias2020photonic}, an analog RMS error of 6$\%$ corresponds to 4 bits, and 0.2$\%$ RMS error corresponds to $\sim$9 bits (see Methods).

We also verified that we could compute dot products between shorter vectors when using low numbers of photons per scalar multiplication (Figure \ref{figure2}c). For photon budgets ranging from 0.001 to 0.1 photons per multiplication, the numerical error was dominated by shot noise for all vector sizes tested. When the number of photons used was sufficiently large, the error was no longer dominated by shot noise, which is consistent with the single-vector-size results shown in Figure \ref{figure2}b. For every photon budget tested, dot products between larger vectors had lower error; we attribute this to dot products between larger vectors involving the effective averaging of larger numbers of terms. 

\section{ONN using sub-photon multiplications}

\begin{figure}[h!]
    \centering
    \includegraphics[width=0.68\textwidth]{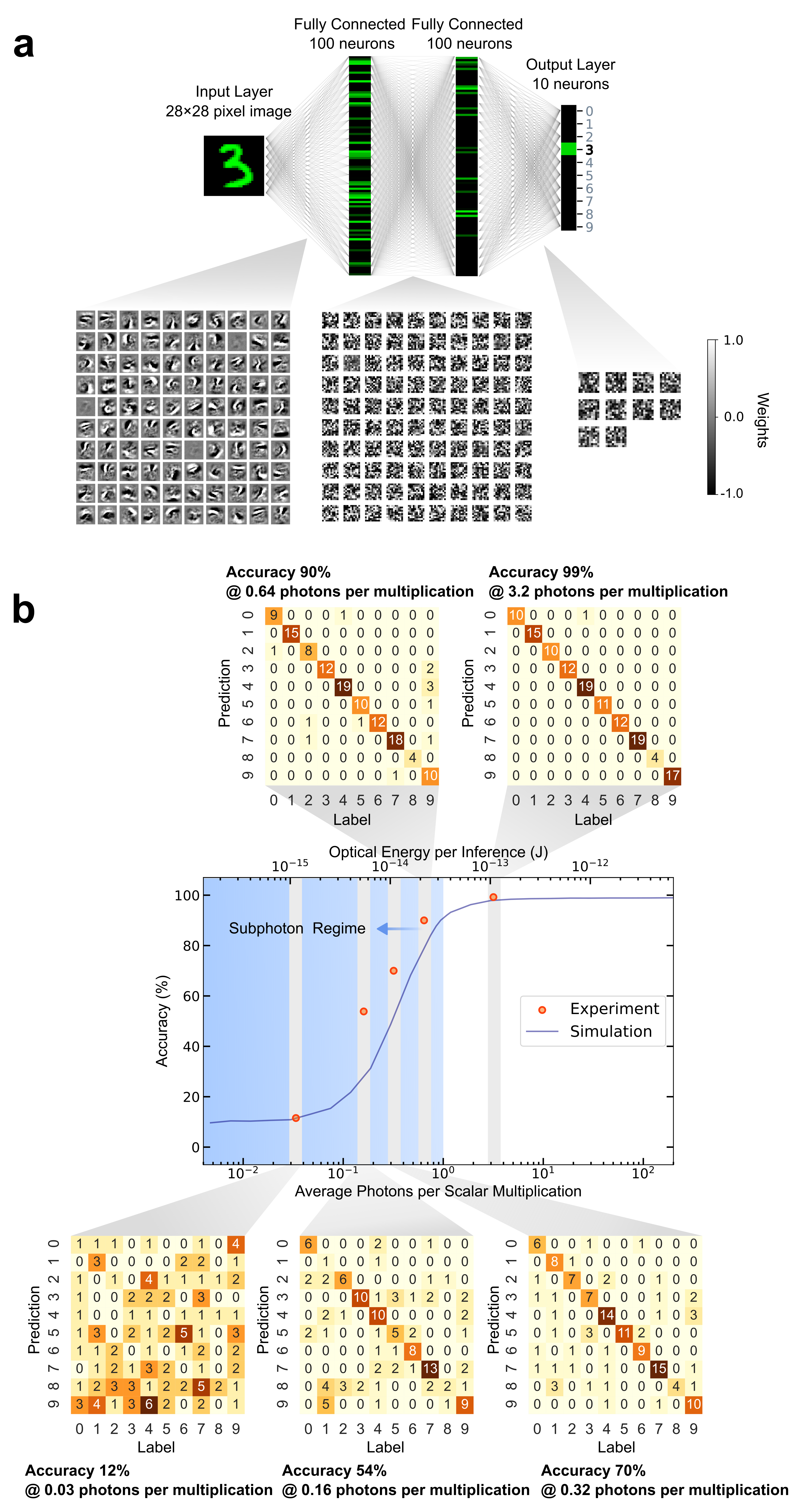}
\end{figure}
\begin{figure}[ht!]
    \caption{\textbf{MNIST handwritten-digit classification demonstrated with an optical neural network using less than one photon per scalar multiplication.} 
    \textbf{a,} Illustration of the neural-network architecture for handwritten-digit classification that we implemented as an ONN. The neural network is composed of a sequence of fully connected layers: an input layer consisting of 28$\times$28 neurons encodes the image to be classified (left); two hidden layers consisting of 100 neurons each (middle), and an output layer containing 10 neurons (right). The activation of each neuron is illustrated as green brightness for the example of an input image containing the digit 3 (vertical bars, top panel). The weights of the connections between neurons for all four layers are shown (grayscale square arrays, bottom panel); each square array shows the weights \textit{from} all the neurons in one layer \textit{to} one of the neurons in the next layer. \textbf{b,} Classification accuracy tested using the MNIST dataset as a function of optical energy consumption (middle panel), and confusion matrices of each corresponding experiment data point (top and bottom panels). The \textit{optical energy per inference} is defined as the total optical energy received by the photodetectors during execution of the three matrix-vector multiplications comprising a single forward pass through the entire neural network.
}
    \label{figure3}
\end{figure}

Having characterized the accuracy of our experimental setup for performing multiplication operations with random vectors, we set out to demonstrate its use as the core of an experimental ONN implementation. We realized an ONN comprised of fully connected layers where the matrix-vector multiplications between each layer were performed optically using our experimental setup, and where the nonlinearity was applied electronically (using a digital processor) between each layer.

Our main goal was to determine the extent to which our ONN could tolerate multiplication inaccuracy resulting from the use of a very limited photon budget. Our approach was to run a trained neural network with our setup and measure the classification accuracy as a function of the number of photons used. We used handwritten-digit classification with the MNIST dataset as our benchmark task and trained a four-layer fully connected multi-layer perceptron (MLP) (Figure \ref{figure3}a) with a back-propagation procedure designed for use with low-precision inference hardware (quantization-aware training—QAT \cite{jacob2017quantization}; see Methods).

We evaluated the first 130 test images in the MNIST dataset under 5 different photon budgets: 0.03, 0.16, 0.32, 0.64, and 3.2 photons per scalar multiplication (Figure \ref{figure3}b, center panel, orange dots). We found that using 3.2 photons per multiplication led to a classification accuracy of $\sim$99$\%$ (Figure \ref{figure3}b, top-right panel), which was almost identical to the accuracy (99$\%$) of the same trained neural network run on a digital computer. In the sub-photon regime, using 0.64 photons per multiplication, the ONN achieved \textgreater90$\%$ classification accuracy (Figure \ref{figure3}b, top-middle panel). The experimental results agree well with the results from simulations of the same neural network being executed by an ONN that is subject to shot noise (Figure \ref{figure3}b, center panel, dark-blue line). To achieve an accuracy of 99$\%$, the total optical energy \textit{detected} for each inference of a handwritten digit was $\sim$\SI{1} {\pico \joule} (Figure \ref{figure3}b). For the weight matrices used in these experiments, the average SLM transmission was $\sim$46$\%$, so when considering the unavoidable loss at the SLM, the total optical energy needed for each inference was $\sim$\SI{2.2} {\pico \joule}. For comparison, $\SI{1}{\pico \joule}$ is close to the energy used for only a single scalar multiplication in electronic processors \cite{jouppi2017datacenter}, and our model required 89,400 scalar multiplications per inference (see Supplementary Information Section 14).

\section{Discussion}
Here we have presented experimental results that support the notion that optical neural networks can in principle \cite{shastri2021photonics,Wetzstein2020,hamerly2019large,nahmias2020photonic} have a fundamental energy advantage over electronic neural-network implementations. We showed that ONNs can operate in a photon-budget regime in which the standard quantum limit (i.e., optical shot noise) governs the achievable accuracy. In particular, we achieved high classification accuracies using our ONN even when restricted to a photon budget of less than one photon per scalar multiplication. We used a standard neural-network model architecture and standard training techniques, so did not specialize the model to our hardware, nor did we need to perform any re-training to run the model on our hardware setup. This successful separation of software and hardware development shows that our ONN could be used as a drop-in replacement for other more conventional neural-network-accelerator hardware \cite{sze2017efficient} without the need for any major changes to the machine-learning software workflow.

Our results have been enabled by several key design choices. The 2D-block design presented in this work, which can be seen as a generalization of the \textit{Stanford matrix-vector multiplier} \cite{goodman1978fully} that avoids cylindrical lenses and their practical limitations, takes full advantage of the number of parallel modes in 3D free space, allowing extremely large numbers of scalar multiplications to be performed in parallel—up to $\sim$0.5 million in our experimental realization. This enabled us to implement a fully parallelized optical matrix-vector multiplier for matrices having size up to 711$\times$711. The same optical setup could be, and was, used to compute vector-vector dot products with vectors of size up to $\sim$0.5 million, which is many orders of magnitude larger than the vector sizes used in previous demonstrations of optical processors \cite{feldmann2021parallel,tait2019silicon,spall2020fully,xu2020single}. The use of 2D blocks was also an important contributor to the achieved accuracy, since this layout reduces the impact of crosstalk between pixels (Supplementary Information Section 10).

We have shown that the \textit{optical} energy used to perform each scalar multiplication can be \textless\SI{1E-18}{\joule} in a functional ONN performing MNIST handwritten-digit classification. The ONN demonstrated in this work was of moderate size, comprising layers with at most 784 neurons, and even better energy efficiency can be achieved for neural networks with wider layers \cite{hamerly2019large}. Our results support an estimate that an ONN with appropriate system-level design could have an advantage of at least two orders-of-magnitude over electronic DNN accelerators, using a total energy across the entire system of only \SI{1E-16}{\joule} per scalar multiplication, even when operating at GHz speeds (Supplementary Information Section 15).

Our proof-of-principle results for sub-photon-per-multiplication ONN operation should readily translate to other ONN architectures, including those using coherent light, provided that they involve summation of a sufficiently large number of modes (be they spatial, frequency, or temporal modes) \cite{hamerly2019large, nahmias2020photonic}.

One critical step towards building practical ONNs with high overall energy efficiency is to design scalable optical fan-out and fan-in using miniaturized passive components. Optical fan-out is crucial to realizing sub-photon scalar multiplications: an optical fan-out stage \cite{goodman1978fully} would be used to spread photons from a single light source across a number of spatial modes larger than the number of photons emitted (in the experiments reported in this paper, less than 1 photon per mode was achieved by applying attenuation, which, while enabling our scientific demonstration, is not suitable as an ultimate engineering solution). Both optical fan-out and optical fan-in can be implemented, for example, with lens arrays \cite{andregg2018wavelength,hayasaki1992optical}; demonstrating an ONN using highly scalable optical fan-out/in stages, integrated with high-efficiency modulators \cite{su2020silicon,burr2015experimental} and detectors \cite{youngblood2015waveguide}, is an important next step. 

More broadly, the ability to perform low-precision matrix-vector multiplications more efficiently could find application beyond neural networks, including in other machine-learning algorithms \cite{de2015taming} and for combinatorial-optimization heuristics \cite{prabhu2020accelerating,mcmahon2016fully,inagaki2016coherent}.

\section*{Data and code availability}
The datasets generated during the characterization of dot-product accuracy (Figure \ref{figure2}) and the execution of the ONN using different photon budgets (Figure \ref{figure3}), along with the code used to analyze them, are available at: \url{https://doi.org/10.5281/zenodo.4722066}. The code we used to train neural networks with QAT in PyTorch is available at: \url{https://github.com/mcmahon-lab/ONN-QAT-SQL}. The code for controlling the devices used in the experiments is available at: \url{https://github.com/mcmahon-lab/ONN-device-control}.

\section*{Acknowledgments}
P.L.M. acknowledges membership of the CIFAR Quantum Information Science Program as an Azrieli Global Scholar. T.W. and L.G.W. acknowledge partial financial support from the Mong Cornell Neurotech Fellow Program. The authors wish to thank NTT Research for their financial and technical support. We thank Frank Wise for the loan of spatial light modulators, and Chris Xu for the loan of imaging lenses that were used in our preliminary experiments. We thank Ryan Hamerly for a helpful discussion on the energy scaling of optical fan-in.

\textit{Competing interests:} T.W. and P.L.M. are listed as inventors on a U.S. provisional patent application (No. 63/149,974) on the techniques to implement 2D-block optical matrix-vector multipliers.

\clearpage
\bibliographystyle{pnas_modified.bst}
\bibliography{references}

\clearpage
\section*{Methods}
\subsection*{Experimental Setup}

We used the OLED display of an Android phone (Google Pixel 2016) as the incoherent light source for encoding input vectors to the matrix-vector multiplier. Only green pixels (with an emission spectrum centered around \SI{525} {\nano \metre}) were used in the experiments; the OLED display contains an array of $\sim$2 million (1,920x1,080) green pixels (Supplementary Information Fig. S3). Custom Android software was developed to load bitmap images onto the OLED display through Python scripts running on a control computer. The phone was found capable of displaying 124 distinct brightness levels ($\sim$7 bits) in a linear brightness ramp (Supplementary Information Fig. S3). At the beginning of each matrix-vector-multiplication computation, the vector was reshaped into a 2D block (Figure \ref{figure1}a) and displayed as an image on the phone screen for the duration of the computation. The brightness of each OLED pixel was set to be proportional to the value of the non-negative vector element it encoded. Fan-out of the vector elements was performed by duplicating the vector block on the OLED display.

Scalar multiplication of vector elements with non-negative numbers was performed by intensity modulation of the light that was emitted from the OLED pixels. An intensity-modulation module was implemented by combining a phase-only reflective liquid-crystal spatial light modulator (SLM, P1920-500-1100-HDMI, Meadowlark) with a polarizing beam splitter and a half-wave plate in a double-pass configuration (Supplementary Information Section 3). An intensity look-up table (LUT) was created to map SLM pixel values to transmission percentages, with an 8-bit resolution (Supplementary Information Section 3).   

Element-wise multiplication between two vectors $\vec{w}$ and $\vec{x}$ was performed by aligning the image of each OLED pixel (encoding an element of $\vec{x}$) to its counterpart pixel on the SLM (encoding an element of $\vec{w}$) (Figure \ref{figure1}b). By implementing such pixel-to-pixel alignment, as opposed to aligning patches of pixels to patches of pixels, we maximized the size of the matrix-vector multiplication that could be performed by this setup. A zoom-lens system (Resolve 4K, Navitar) was employed to de-magnify the image of the OLED pixels by $\sim$0.16$\times$ to match the pixel pitch of the SLM (Supplementary Information Section 5). The image of each OLED pixel was diffraction-limited with a spot diameter of $\sim$\SI{6.5} {\micro \metre}, which is smaller than the \SI{9.2} {\micro \metre} size of pixels in the SLM, to avoid cross-talk between neighboring pixels. Pixel-to-pixel alignment was achieved for $\sim$0.5 million pixels (Supplementary Information Section 5). This enabled the setup to perform vector-vector dot products with 0.5-million-dimensional vectors in single passes of light through the setup (Figure \ref{figure2}b). The optical fan-in operation was performed by focusing the modulated light field onto a detector, through a 4f system consisting of the rear adapter of the zoom-lens system and an objective lens (XLFLUOR4x/340, NA=0.28, Olympus) (Fig. S1 and Supplementary Information Section 9).

The detectors measured optical power by integrating the photon flux incident on the detector’s active area over a specified time window. Different types of detector were employed for different experiments. A multi-pixel photon counter (MPPC, C13366-3050GA, Hamamatsu) was used as a bucket detector for low-light-level measurements. This detector has a large dynamic range (\SI{}{\pico\watt} to \SI{}{\nano\watt}) and moderately high bandwidth ($\sim$\SI{3}{\mega \hertz}) (Supplementary Information Section 4). The MPPC outputted a single voltage signal representing the integrated optical energy of the spatial modes focused onto the detector area by the optical fan-in operation (Supplementary Information Section 9). The MPPC is capable of resolving the arrival time of single-photon events for low photon fluxes (\SI{<e6}{} per second); for higher fluxes that exceed the bandwidth of MPPC ($\sim$\SI{3}{\mega \hertz}), the MPPC output voltage is proportional to the instantaneous optical power (Fig. S6 and Supplementary Information Section 4). The SNR of the measurement of optical energy with MPPC is about half of the shot-noise-limited SNR (Supplementary Information Section 8). Since the MPPC does not provide spatial resolution within its active area, it effectively acts as a single-pixel detector (Figure 1c) and consequently could only be used to read out one dot product at a time. For parallel computation of multiple dot products (as is desirable when performing matrix-vector multiplications that are decomposed into many vector-vector dot products), a CMOS camera (Moment-95B, monochromatic, Teledyne) was used. The intensity of the modulated light field was captured by the camera as an image, which was divided into regions of interest (ROIs), each representing the result of an element-wise product of two vectors. The pixels in each ROI could be then summed digitally to obtain the total photon counts, which correspond to the value of the dot product between the two vectors. Compared to the MPPC, the CMOS camera was able to capture the spatial distribution of the modulated light but could not be used for the low-photon-budget experiments due to its much higher readout noise ($\sim$2 electrons per pixel) and long frame-exposure time ($\geq$\SI{10}{\micro \second}). Consequently the camera was only used for setup alignment and for visualizing the element-wise products of two vectors with large optical powers, and the MPPC was used for the principal experiments in this work—namely, vector-vector dot-product calculation and matrix-vector multiplication involving low numbers of photons per scalar multiplication (Figure \ref{figure2} and Figure \ref{figure3}).

\subsection*{Evaluation of Dot-Product Accuracy}

The numerical accuracy of dot products was characterized with pairs of vectors consisting of non-negative elements; since there is a straightforward procedural modification to handle vectors whose elements are signed numbers, the results obtained are general (Supplementary Information Section 11). The dot-product answers were normalized such that the answers for all the vector pairs used fall between 0 and 1; this normalization was performed such that the difference between true and measured answers could be interpreted as the achievable accuracy in comparison to the full dynamic range of possible answers (for the equations used for the error calculation, see Supplementary Information Section 12).

Before the accuracy-characterization experiments were performed, the setup was calibrated by recording the output of the detector for many different pairs of input vectors and fitting the linear relationship between the dot-product answer and the detector’s output (Supplementary Information Section 12). 

The vector pairs used for accuracy characterization were generated from randomly chosen grayscale natural-scene images (STL-10 dataset \cite{coates2011analysis}). The error of each computed dot product was defined as the difference between the measured dot-product result and the ground truth calculated by a digital computer (Figure \ref{figure2}a). The number of photons detected for each dot product was tuned by controlling the integration time window of the detector (Supplementary Information Section 12). The measurements were repeated many times to capture the error distribution resulting from noise. For each vector size displayed in Figure \ref{figure2}c, the dot products for 100 vector pairs were computed. The root-mean-square (RMS) error was calculated based on data collected for different vector pairs and multiple measurement trials. Therefore, the RMS error includes contributions from both the systematic error and trial-to-trial error resulting from noise. The RMS error can be interpreted as the “expected” error from a single-shot computation of a dot product with the optical processor. The noise equivalent bits were calculated using the formula \cite{nahmias2020photonic} $\text{NEB}=-\log_2(\text{RMS Error}$).

\subsection*{Training of Noise-resilient Neural Networks}

To perform handwritten-digit classification, we trained a neural network with 4 fully connected layers (Figure \ref{figure3}a). The input layer consists of 784 neurons, corresponding to the $28 \times 28 =784$ pixels in grayscale images of handwritten digits. This is followed by two fully connected hidden layers with 100 neurons each. We used ReLU \cite{glorot2011deep} as the nonlinear activation function. The output layer has 10 neurons; each neuron corresponds to a digit from 0 to 9, and the prediction of which digit is contained in the input image is made based on which of the output neurons had the largest value. The neural network was implemented and trained in PyTorch \cite{paszke2019pytorch}. The training of the neural network was conducted exclusively on a digital computer (our optical experiments perform neural-network inference only). To improve the robustness of the model against numerical error, we employed quantization-aware training (QAT) \cite{jacob2017quantization}, which was set to quantize the activations of neurons to 4 bits and weights to 5 bits. The PyTorch implementation of QAT was adapted from Ref. \cite{hubara2016binarized}. In addition, we performed data augmentation: we applied small random affine transformations and convolutions to the input images during training. This is a standard technique in neural-network training for image-classification tasks to avoid overfitting \cite{shorten2019survey} and intuitively should also improve the model’s tolerance to potential hardware imperfections (e.g., image distortion and blurring). The training parameters we used are documented in Supplementary Information Section 13. The training methods used not only effectively improved model robustness against numerical errors but also helped to reduce the optical energy consumption during inference. We note that the 4-bit quantization of neuron activations was only performed during training, and not during the inference experiments conducted with the optical matrix-vector multiplier: the activations were loaded onto the OLED display using the full available precision (7 bits).

\subsection*{Optical Neural Networks with Controlled Photon Budgets}

To execute the trained neural network with the optical matrix-vector multiplier, we needed to perform 3 different matrix-vector multiplications, each responsible for the forward propagation from one layer to the next. The weights of each matrix of the MLP model were loaded onto the SLM, and the vector encoding the neuron values for a particular layer was loaded onto the OLED display. (There is a technicality associated with the handling of negative values, as was mentioned in the above Methods section on dot-product characterization and is explained in detail in Supplementary Information Section 11.) We performed matrix-vector multiplication as a set of vector-vector dot products. For each vector-vector dot product, the total photon counts (or optical energy) measured by the detector were mapped to the answer of the dot product through a predetermined calibration curve. The calibration curve was made using the first 10 samples of the MNIST test dataset by fitting the measured photon counts to the ground truth of the dot products (Supplementary Information Section 14). The number of photons per multiplication was controlled by adjusting the detector’s integration time (Supplementary Information Section 12). The measured dot-product results were communicated to a digital computer where bias terms were added and the nonlinear activation function (ReLU) was applied. The resulting neuron activations of each hidden layer were used as the input vector to the matrix-vector multiplication for the next weight matrix. At the output layer, the prediction was made in a digital computer based on the neuron with the highest value.

\end{document}

Supplementary Information for 
\title{An optical neural network using less than 1 photon per multiplication}

\author{Tianyu~Wang}
\email{tw329@cornell.edu}
\affiliation{School of Applied and Engineering Physics, Cornell University, Ithaca, NY 14853, USA}

\author{Shi-Yuan~Ma}
\affiliation{School of Applied and Engineering Physics, Cornell University, Ithaca, NY 14853, USA}

\author{Logan~G.~Wright} 
\affiliation{School of Applied and Engineering Physics, Cornell University, Ithaca, NY 14853, USA}
\affiliation{NTT Physics and Informatics Laboratories, NTT Research, Inc., Sunnyvale, CA 94085, USA}

\author{Tatsuhiro~Onodera}
\affiliation{School of Applied and Engineering Physics, Cornell University, Ithaca, NY 14853, USA}
\affiliation{NTT Physics and Informatics Laboratories, NTT Research, Inc., Sunnyvale, CA 94085, USA}

\author{Brian~C.~Richard}
\affiliation{School of Electrical and Computer Engineering, Cornell University, Ithaca, NY 14853, USA}

\author{Peter~L.~McMahon}
\email{pmcmahon@cornell.edu}
\affiliation{School of Applied and Engineering Physics, Cornell University, Ithaca, NY 14853, USA}

\maketitle

\tableofcontents
\clearpage

\part{Experimental Setup}

\section{Overview of the Experimental Setup and Components} \label{overview_setup}

\begin{figure}[ht!]
\includegraphics [width=0.78\textwidth] {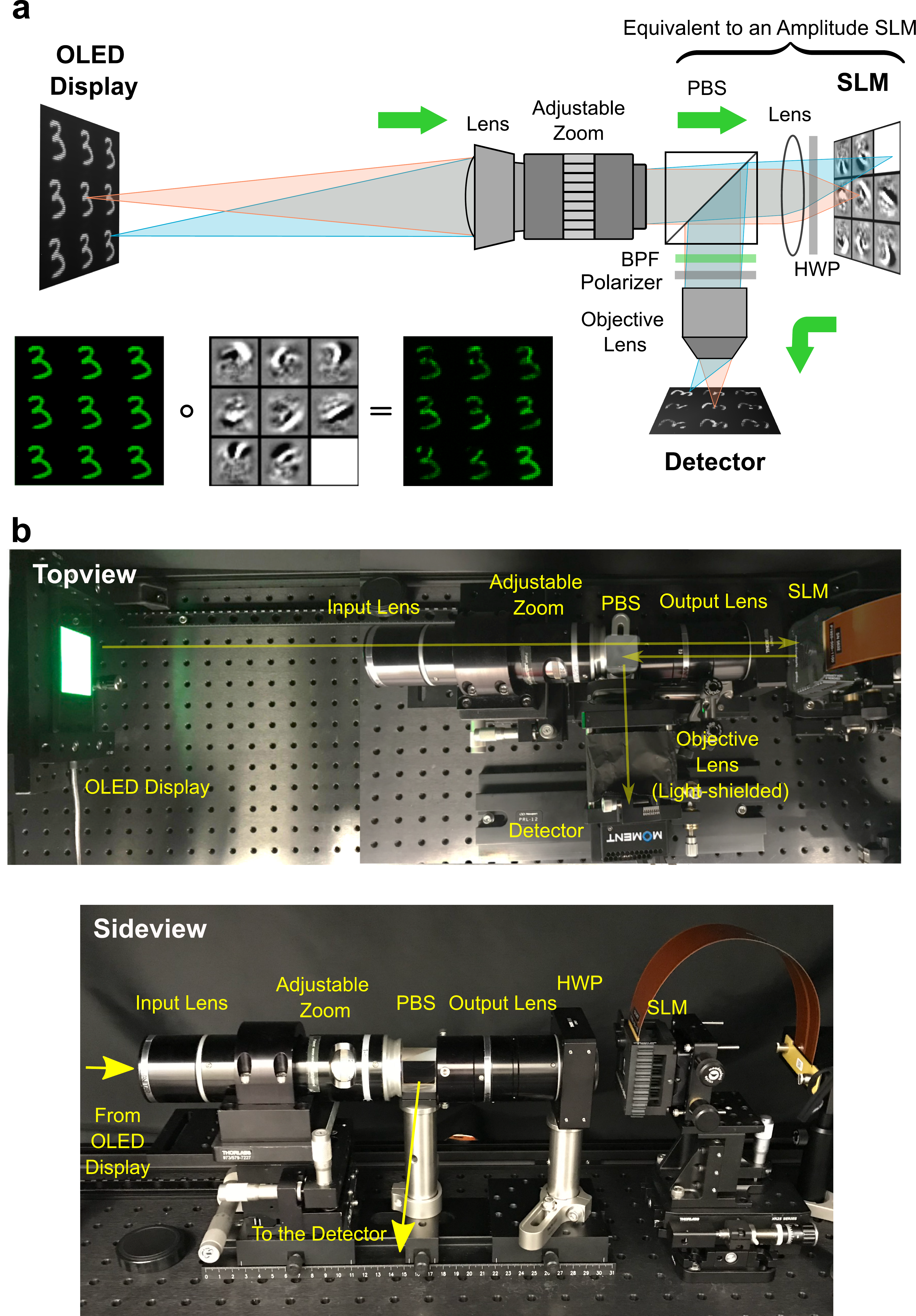}
\caption{\textbf{Experimental Layout for the Optical Matrix-vector Multiplier Setup.}  \textbf{a,} The schematic of the optical setup. An illustration of the element-wise multiplication is also shown. In this example, nine copies of an input vector of handwritten digits, all '3'---which our setup accepts as 2D images---are intensity modulated by different weight vectors, which are each encoded as a 2D block on the SLM. Images of the vectors before and after the intensity modulation were taken by a camera placed at the detector location. (PBS: polarizing beam splitter; HWP: half-wave plate; BPF: band-pass filter) \textbf{b,} Photos of the core setup corresponding to the schematic are shown in panel (a). }
\label{setup_schematics}
\end{figure}

\begin{figure}[ht!]
\includegraphics [width=0.9\textwidth] {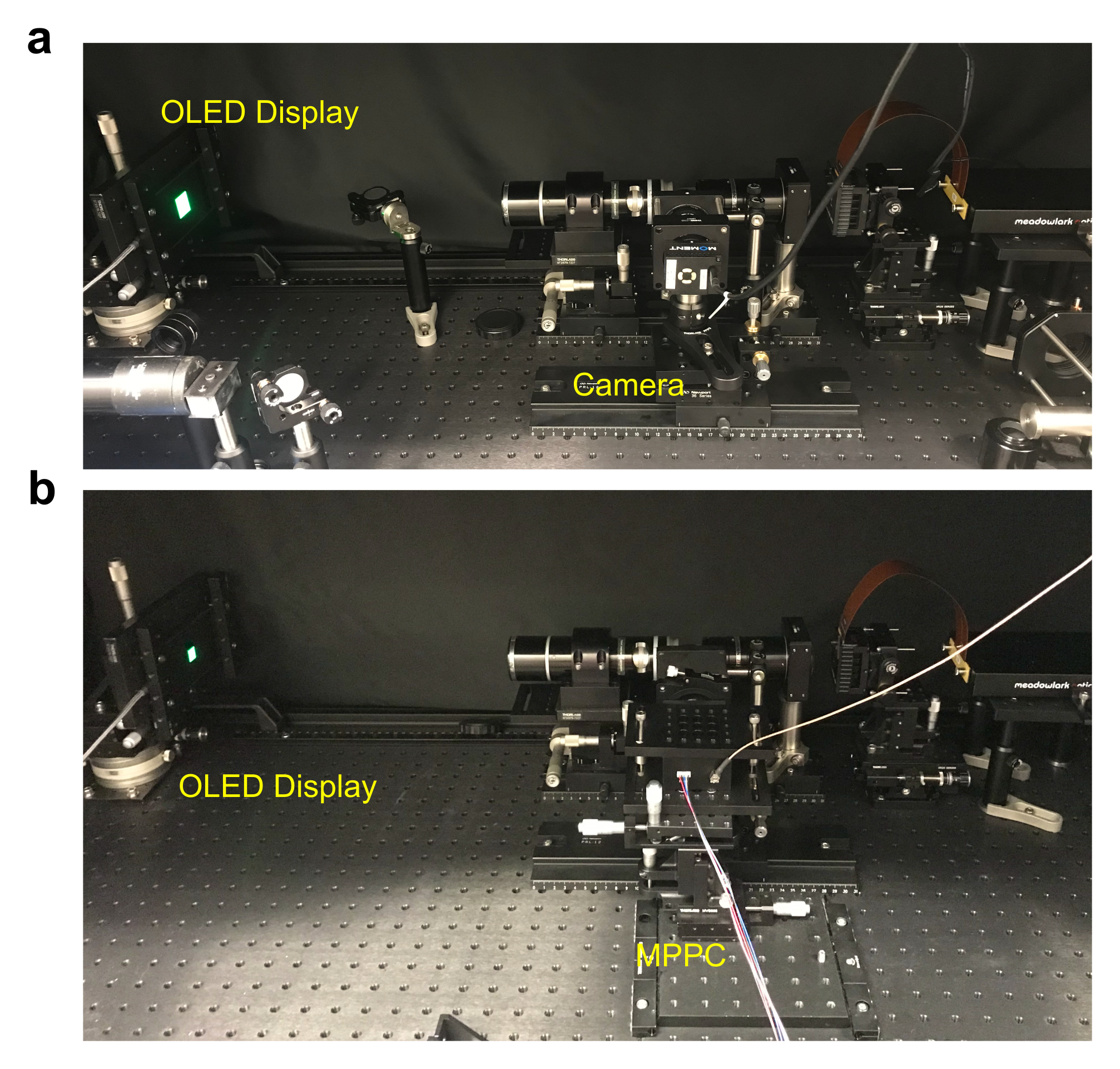}
\caption{\textbf{Photos of the Entire Setup.} \textbf{a,} The setup configured with a CMOS camera as a multi-pixel detector. \textbf{b,} The setup configured with an MPPC as a sensitive bucket detector.}
\label{setup}
\end{figure}

The optical matrix-vector multiplier setup consists of an array of light sources, a zoom lens imaging system, an intensity modulator, and a photodetector (Fig. \ref{setup_schematics}). We used an organic light-emitting diode (OLED) display of a commercial smartphone (Google Pixel 2016 version) as the light source for encoding input vectors. The OLED display consist of a $1920 \times 1080$ pixel array, with individually controllable intensity for each pixel (for details, see Section \ref{OLED_sec}). A reflective liquid-crystal spatial light modulator (SLM, P1920-500-1100-HDMI, Meadowlark Optics) was combined with a half-wave plate (HWP, WPH10ME-532, Thorlabs) and a polarizing beamsplitter (PBS, CCM1-PBS251, Thorlabs) to perform intensity modulation as weight multiplication (for details, see Section \ref{SLM_mod}). The SLM has a pixel array of dimensions $1920 \times 1152$, with individually controllable transmission for each pixel. A zoom lens system (Resolv4K, Navitar) was used to image the OLED display onto the SLM panel (for details, see Section \ref{alignment}). The intensity-modulated light field reflected from the SLM was further de-magnified and imaged onto the detector, by a telescope formed by the rear adapter of the zoom lens (1-81102, Navitar) and an objective lens (XLFLUOR4x/340, Olympus). An additional band-pass filter (BPF, FF01-525/15-25, Semrock) and polarizer (LPVISE100-A, Thorlabs) were inserted into the telescope (Fig. \ref{setup_schematics}) in order to reduce the bandwidth and purify the polarization of the light reflected by the PBS, resulting in more precise results. During alignment and troubleshooting, we used a camera (Prime 95B Scientific CMOS Camera, Teledyne Photometrics) as a multi-pixel detector (Fig. \ref{setup}a). For sensitive measurements under extremely low photon fluxes, we used a multi-pixel photon counter (MPPC, C13366 series GA type, Hamamatsu Photonics) as a bucket detector (Fig. \ref{setup}b) (for details, see Section \ref{optical_fan_in}). When it was necessary to further reduce the optical power, an additional neutral density filter ($\text{ND=0.4}$, NE2R04B, Thorlabs) was placed in front of the zoom lens to attenuate light.

\clearpage
\section{Properties of the OLED Display} \label{OLED_sec}

We chose an OLED display as the light source for several reasons. Unlike coherent sources (e.g., lasers), OLEDs are unable to encode phase information and generally have slower modulation speeds when used as an incoherent light source. However, the technology to integrate millions of OLED pixels is more commercially available and is less expensive than integrated laser units (e.g., VCSELs) or passive modulator arrays. A high pixel count enabled us to encode very large vectors, and was essential for demonstrating the optical energy advantage of our setup. 

\begin{figure}[ht!]
\includegraphics [width=\textwidth] {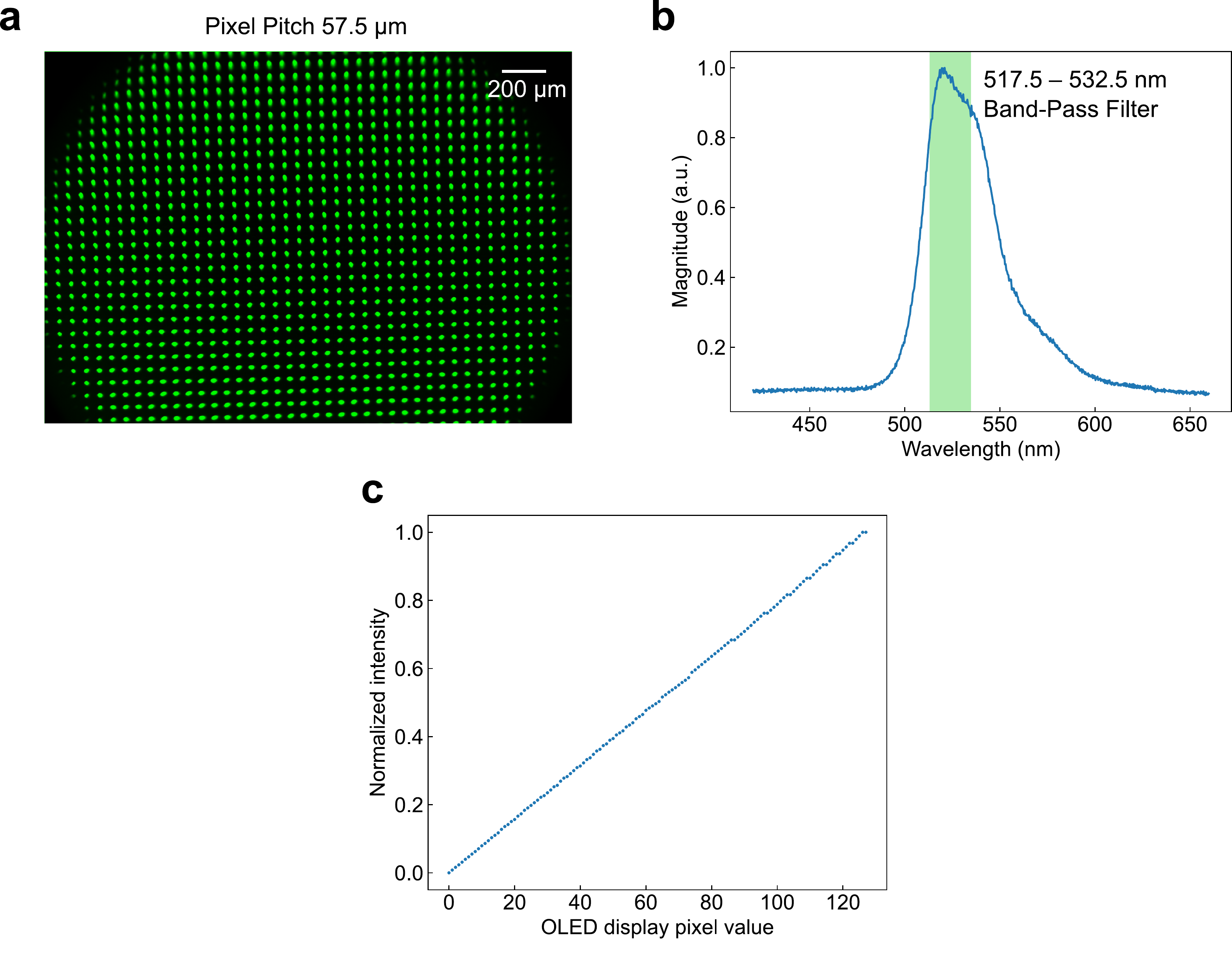}
\caption{\textbf{Properties of the Organic Light-Emitting Diode (OLED) Display.} \textbf{a,} An image of the green OLED pixels taken under an inspection microscope. The pixels form a square lattice with a pixel pitch of \SI{57.5}{\micro\metre}. Scale bar, \SI{200}{\micro\metre}. \textbf{b,} Emission spectrum of the green OLED pixels. The shaded area indicates the transmission band of the BPF ($>90\%$ transmission). \textbf{c, }The 7-bit linear look-up table (LUT) calibrated to control the OLED display intensity.
}
\label{OLED}
\end{figure}

Compared to other options of integrated incoherent light sources (e.g., liquid-crystal display, LCD) OLED pixels can be turned off completely, while LCD screens are always backlit with LED panels and thus always transmit some residual light. The true darkness of OLED pixels allowed us to achieve high dynamic range in intensity modulation and to reduce noise caused by background light pollution. Finally, the OLED display pixels used in this experiment were conveniently point-shaped and arranged in the same square lattice array as the SLM pixels (Fig. \ref{OLED}a), which facilitated pixel-to-pixel alignment (see Section \ref{alignment}). In contrast, commercial LCD pixels are typically arranged in bars, which require optical image transformation before they can be aligned to SLM pixels.

For our study, we used an OLED display with three different colors of pixels: red, blue, and green. We used only the green pixels, which form a $1080 \times 1920$ square lattice array (\SI{\sim 2e6}{} total pixels) as shown in Fig. \ref{OLED}a. The pixel pitch was measured to be \SI{57.5}{\micro\metre}. The maximum power of each pixel was measured to be \SI{\sim 1}{\nano\watt}, emitted in a very wide angle ($> 60$ degrees). Since the light emitted from the OLED screen had a rather broad spectrum, we used a band-pass filter (FF01-525/15-25, Semrock) to reduce the bandwidth in order to improve coherence for more precise and stable phase modulation by the SLM (Fig. \ref{OLED}b). The intensity of each individual pixel could be controlled independently with 256 (8-bit) control levels. However, since the actual output intensity was not linear with the pixel control level, we calibrated a linear look-up table (LUT) that contains 124 distinct intensity levels ($\sim$7 bits, Fig. \ref{OLED}c).

\section{Intensity Modulation with a Phase-only SLM} \label{SLM_mod}

\begin{figure}[h!]
\includegraphics [width=0.9\textwidth] {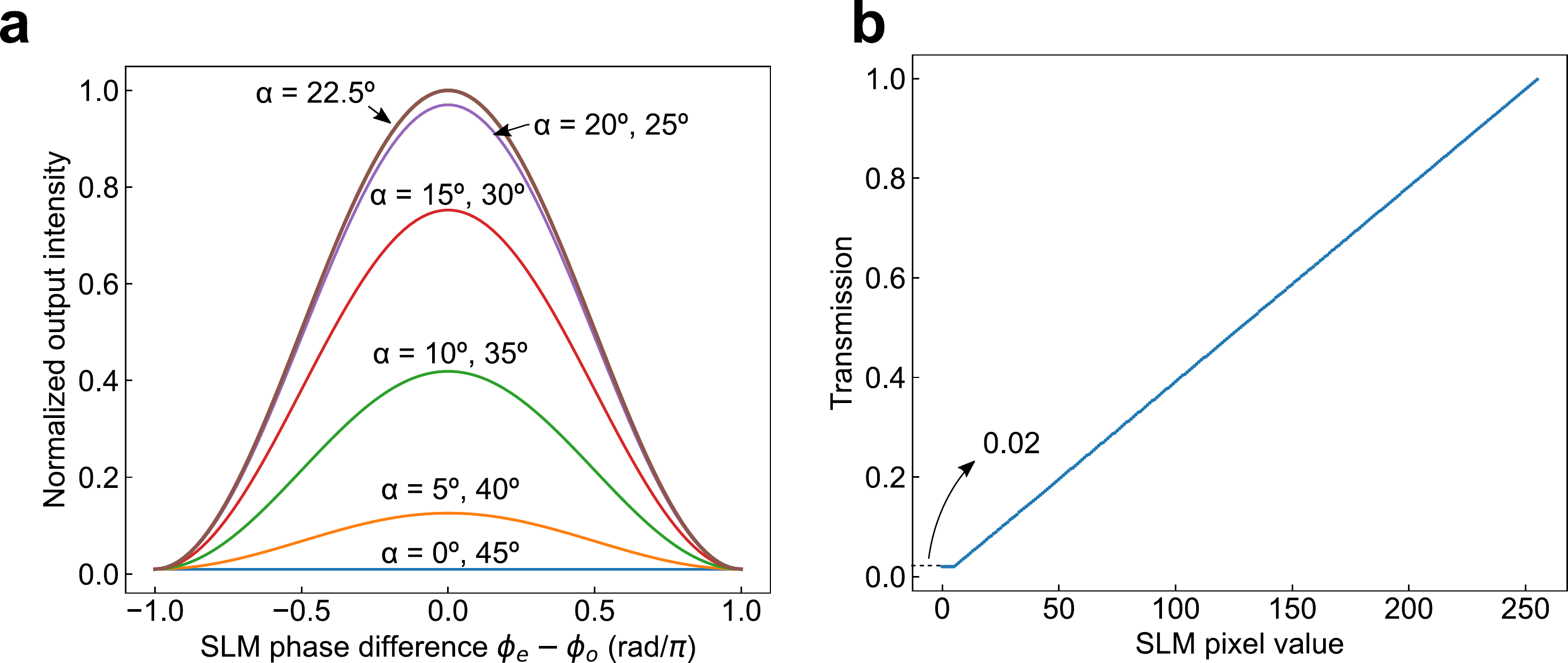}
\caption{\textbf{ Intensity Modulation with a Spatial Light Modulator (SLM).} \textbf{a,} Simulation results of output intensity as a function of phase difference $\phi_e-\phi_o$. $\alpha$ is the angle between the half-wave plate (HWP) fast axis and the extraordinary axis of the SLM. \textbf{b,} The 8-bit LUT of the SLM for intensity modulation. The minimum transmission was measured to be $\sim$0.02 times the maximum transmission, which is equivalent to an extinction ratio of $\sim$50.}
\label{amp_mod}
\end{figure}

We converted a phase-only SLM into an intensity modulator with a half-wave plate (HWP) and a polarizing beam splitter (PBS). The SLM pixels are made of birefringent liquid crystal layers, whose refractive index can be tuned by applying voltage across them. By controlling the refractive index of extraordinary light, the SLM pixels introduce a phase difference $\phi_e-\phi_o$ between the extraordinary and ordinary light, whose polarizations are perpendicular to each other. When a PBS and HWP were placed in front of a reflective SLM, the light field passed the components twice, once during the trip towards the SLM and once after being reflected by the SLM (Fig. \ref{setup_schematics}a). One of the functions of PBS was to separate the output from the input light: the input light (incident to the SLM) was horizontally polarized and transmitted by the PBS, while the output light (reflected from the SLM) was vertically polarized, and therefore reflected by the PBS. The other function of the PBS is to convert the polarization state of the output light to its amplitude: the light modulated by the SLM was in general elliptically polarized, controlled by the phase difference $\phi_e-\phi_o$. The amplitude of the light field (and intensity in this case too) was modulated by selecting only the vertical component of the SLM-modulated light at the output port of the PBS. The HWP was placed with its fast axis rotated 22.5 degrees from the extraordinary axis of the SLM such that the intensity transmission could be tuned from 0 to 100\%. Fig. \ref{amp_mod}a shows the calculated relationship between the intensity transmission and phase difference $\phi_e-\phi_o$. The maximum extinction ratio of the transmission intensity was measured to be $\sim$50 (Fig. \ref{amp_mod}b). The SLM consists of $1920 \times 1152 \sim 2.2 \times 10^6$ pixels, each of which can be independently controlled for intensity modulation with a 256 (8-bit) LUT (Fig. \ref{amp_mod}c). Alternatively, instead of using a phase-modulation SLM, the intensity modulator can be more compactly implemented with a monolithic LCD panel in a transmission geometry. 

\section{Characterization of the Photodetector (MPPC)} \label{MPPC}

\begin{figure}[h!]
\includegraphics [width=0.9\textwidth] {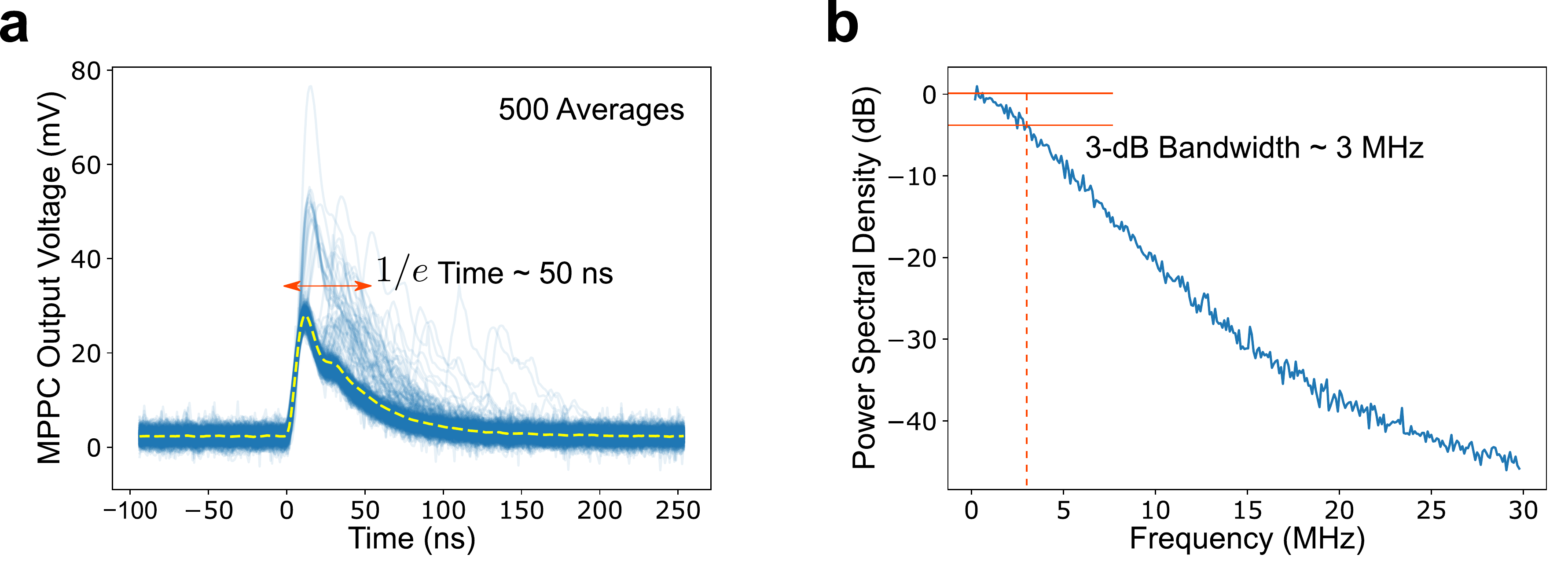}
\caption{\textbf{Time and Frequency Characteristics of the Multi-Pixel Photon Counter (MPPC).} \textbf{a,} The impulse response of single-photon detection averaged over 500 trials. \textbf{b,} The frequency response and Power Spectral Density of the MPPC. }
\label{MPPC_time_char}
\end{figure}

For single-photon detection, we used a multi-pixel photon counter (MPPC) as a bucket detector. We chose an MPPC for its high signal-to-noise ratio (SNR), large measurement range, and moderately high bandwidth. The MPPC is composed of an array of Geiger-mode photodiodes with high intrinsic gain, which enables the photodiodes to detect single-photon events. The detection of each photon results in a spike-shaped impulse response in the output voltage of the detector, with a sharp rising edge and an approximately exponentially decaying tail (Fig. \ref{MPPC_time_char}a). 

\begin{figure}[h!]
\includegraphics [width=\textwidth] {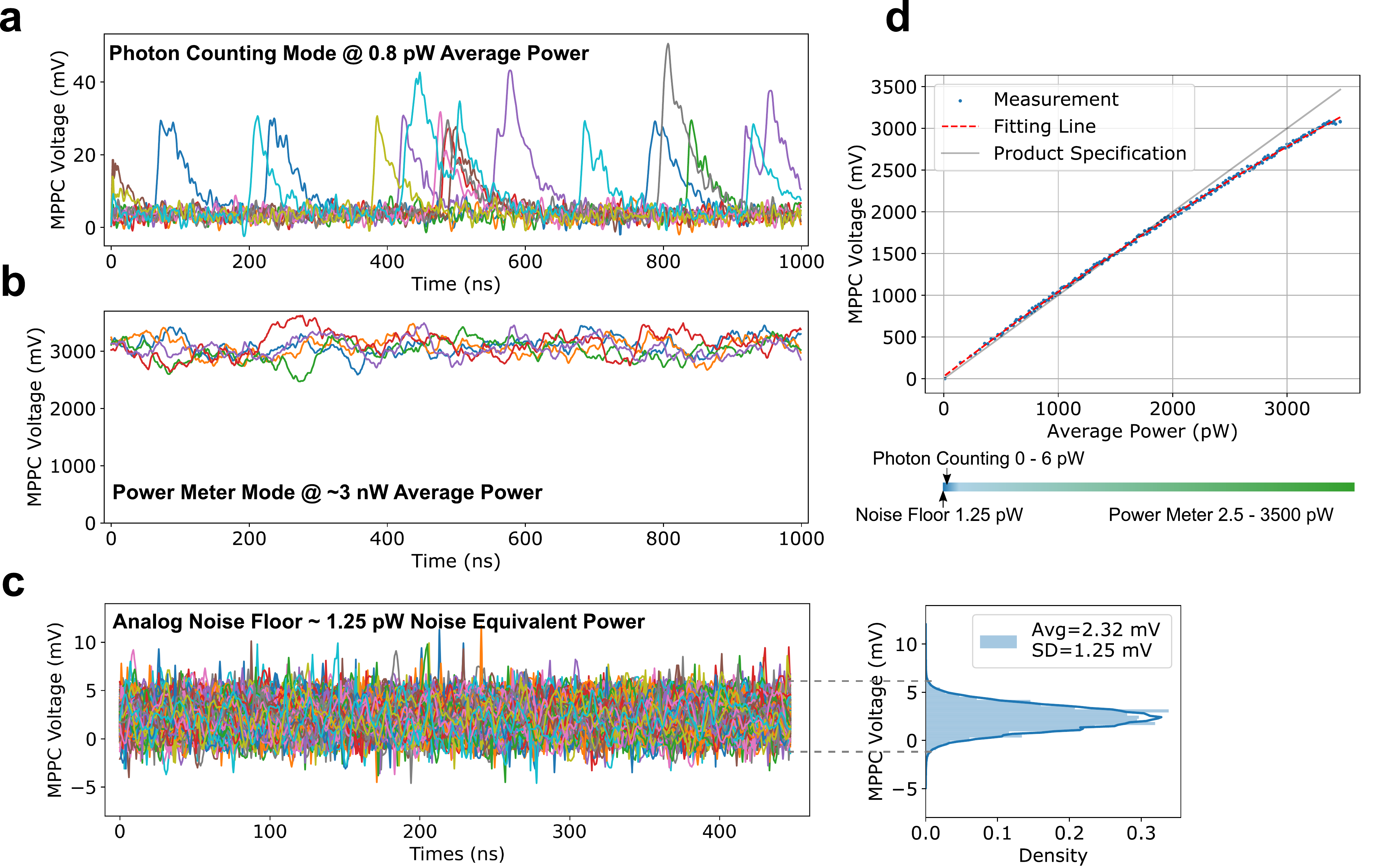}
\caption{\textbf{Photon Detection and Optical Power Measurement with the MPPC.} \textbf{a, }Single-photon detection under low photon flux. Different colors indicate instances of independent measurement trials. \textbf{b, }Instantaneous optical power measurement under high photon flux. \textbf{c, }The detector noise floor, which determines the lowest measurable optical power when the MPPC is used as a power meter. The left panel shows examples of independent measurements of the baseline analog noise; the right panel shows the noise distribution and statistics. \textbf{d, }The linear relationship between the average MPPC output voltage and the average optical power impinging on the detector. Calibration of the optical power (plot group ``Measurement") was performed independently with a semiconductor power meter. }
\label{MPPC_power}
\end{figure}

When the photon flux rate is extremely low (\textless$10^7$ photons per second), the detected photons can be enumerated by counting the number of spikes (Fig. \ref{MPPC_power}a). The maximum measurable photon flux rate is limited by the bandwidth of the detector (Fig. \ref{MPPC_time_char}b) and potentially the dead time after each photon detection. To increase the maximum measurable photon flux (or optical power), the MPPC detector was spatially multiplexed with a $60 \times 60$ photodiode array (with \SI{50} {\micro\metre}$\times$\SI{50} {\micro\metre} pixel pitch), the outputs of which are then pooled into a single analog voltage trace. When the photon flux far exceeds the MPPC bandwidth (\SI{\gg e7}{} photons per second), the pulses induced by individual photons overlap in time and can no longer be resolved (Fig. \ref{MPPC_power}b). In this scenario, we measured the average output voltage, which maintains an excellent linear relationship with the average optical power impinging on the detector. The MPPC output voltage was calibrated against the power reading of a semiconductor power meter (818-UV-L-FC/DB, Newport), and the calibration result closely agreed with the manufacturer’s specifications of the MPPC (Fig. \ref{MPPC_power}d). Therefore, we were able to use the detector as a fast power meter to measure instantaneous optical power from \SI{}{\pico\watt} up to several \SI{}{\nano\watt} (Fig. \ref{MPPC_power}d). In this experiment, optical power \SI{>6} {\pico\watt} was measured by converting the output voltage of the MPPC to the optical power impinging on the detector. Compared to regular semiconductor power meters without intrinsic gain---which can also measure $\sim$\SI{}{\pico\watt} levels of optical power---the MPPC can maintain a high SNR for a much higher bandwidth (\SI{\sim 3} {\mega\hertz}, Fig. \ref{MPPC_time_char}b), since its signal is amplified to overcome the noise integrated over the larger bandwidth. 

When the MPPC is used as a power meter, the minimum power that can be measured is determined by the analog noise floor (including dark counts, thermal noise, and other electronic noises), which was measured to be equivalent to \SI{1.25}{\pico\watt} optical power at the full bandwidth (Fig. \ref{MPPC_power}c). The dark count of the MPPC was measured to be \SI{\sim e4}{} photons per second (\SI{<10}{\femto\watt}), which accounted for less than 1\% of the total noise. Therefore, the detector can in principle measure optical power even below the analog noise equivalent power of \SI{1.25}{\pico\watt} by means of photon counting. In our experiments, the photon counting measurement was conducted only to verify that the detector could indeed resolve single-photon events, and to determine the minimum valid optical power it could measure (\SI{\sim 10} {\femto\watt}). Since the optical powers involved in our experiments were higher than the analog noise floor (\SI{\sim 1.25} {\pico\watt}), they were all measured via the direct readout of detector's output voltage.

\clearpage
\section{Alignment of the Optical Imaging System}\label{alignment}

\begin{figure}[h!]
\includegraphics [width=\textwidth] {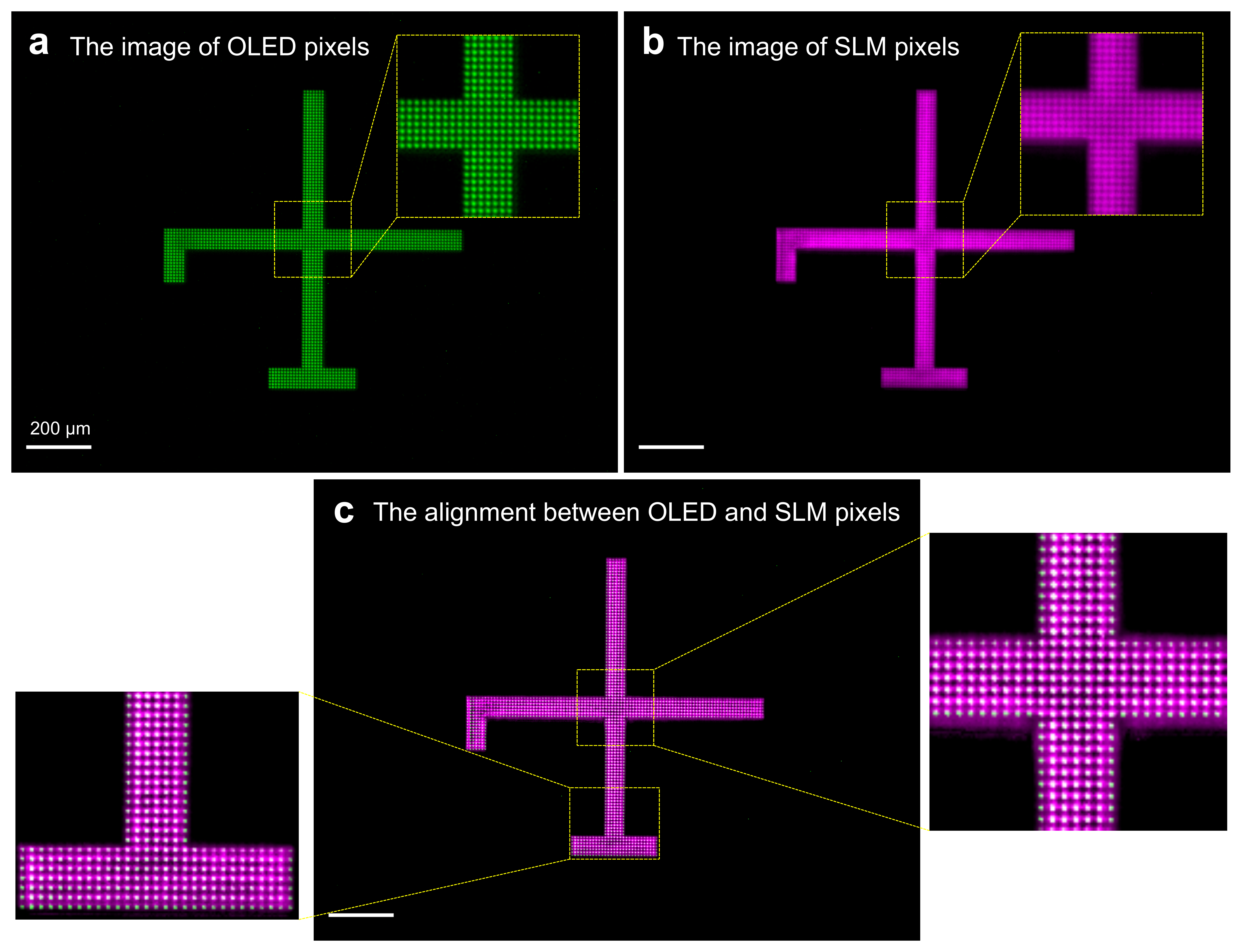}
\caption{\textbf{Pixel-to-pixel Alignment Between the OLED and SLM Pixels.} \textbf{a,} The image of a viewfinder pattern on the OLED display. The scale bar measures the distance on the SLM panel, \SI{200} {\micro\metre}. \textbf{b,} The image of the identical viewfinder pattern modulated by the SLM. The entire SLM panel was uniformly illuminated by an ambient light source. The SLM performed intensity modulation which functioned as a mask with a hollow of the viewfinder shape. \textbf{c,} Visualization of the alignment between the OLED and SLM pixels. }
\label{view_finder}
\end{figure}

In order to maximize the matrix-vector multiplication size---and thus maximize the energy benefits of optical processing---we aligned as many pixels as possible from the OLED display to the SLM. Three conditions must be satisfied for this pixel-to-pixel alignment:
\begin{enumerate}
\item The OLED display must be imaged onto the SLM with a precise de-magnification factor to match the pitch of OLED pixels to that of the SLM pixels.
\item The imaging resolution needs to be high enough such that the image of each OLED pixel on the SLM must be no larger than the size of an SLM pixel. This is to prevent crosstalk.
\item The image of each OLED pixel must be aligned to the corresponding pixel on the SLM, which requires fine adjustment of the translation, rotation, pitch, and yaw of each device involved. 
\end{enumerate}
To match the pixel size of the OLED pixel image to the SLM pixel size, the zoom lens was set at a de-magnification of \SI{9.2} {\micro\metre}$/$\SI{57.5} {\micro\metre} $= 0.16$. This zoom factor was achieved when the zoom lens (Resolv4K 1-80100, Navitar) was configured with a $0.25\times$ lens attachment (1-81201, Navitar) and $1\times$ rear adapter (1-81102, Navitar). The zoom factor of the zoom lens could be mechanically tuned continuously to precisely match the OLED and SLM pixel pitch. Under this configuration, the spot size in the object plane of the zoom lens (the OLED side) is \SI{40.85} {\micro\metre} in diameter (Rayleigh criterion) according to the manufacturer’s  specifications. This spot size is smaller than the OLED pixel pitch size of \SI{57.5} {\micro\metre}. Meanwhile, the spot size in the imaging plane (the SLM side) is specified to be \SI{6.52} {\micro\metre} in diameter (Rayleigh criterion), which is also smaller than the SLM pixel pitch size of \SI{9.2} {\micro\metre}. In fact, the performance of the zoom lens system was close to the diffraction limit, and the images of OLED pixels on the SLM plane were well separated (Fig. \ref{view_finder}). Therefore, the setup achieved the correct de-magnification factor and possessed adequate resolution, and both conditions (1) and (2) were satisfied.

To align each OLED pixel to the corresponding SLM pixel, mechanical alignment was performed using the following method (Fig. \ref{view_finder}). First, identical images of the same size were displayed on both the OLED display and SLM. The bright pixels on the OLED display corresponded to pixels of full transmission on the SLM. The dark pixels on the OLED display corresponded to the pixels of zero transmission on the SLM. Therefore, the SLM functioned like a mask, with its light-transmitting parts identical in shape and size to the bright image on the OLED display. After intensity modulation by the SLM, the image on the OLED can only be preserved without any clipping if and only if the OLED and SLM pixels are exactly aligned to each other, and with the correct orientation (Fig. \ref{view_finder}).

The maximum number of pixels that could be aligned was determined by the imaging error on the side of the field-of-view (FOV) of the zoom lens. According to the specifications of the zoom lens, at most 3.6 million OLED pixels in a square array can be aligned to the SLM. However, this estimation assumes diffraction-limited performance across the entire FOV. In practice, we managed to align $711 \times 711 \sim 0.5$ million pixels. There were three reasons for the deterioration of pixel-to-pixel alignment towards the side of the FOV: 
\begin{enumerate}
\item Vignette: The optical transmission drops off towards the edge of the FOV, which causes up to a \SI{\sim 90}{\percent} decrease in intensity for the $711 \times 711$ pixel array. As a result, the pixels around the center of the FOV must be dimmed in order to keep all pixels at the same brightness, as required for the computation of large dot products.
\item Image Distortion: nonlinear image distortion (such as barrel, or pin cushion, or other higher-order distortions) lead to a non-uniform local zoom factor of the OLED pixel image. Although linear image distortions can be corrected by mechanical alignment, nonlinear distortions cannot be completely fixed by alignment. Even slight distortions can cause pixels to slowly drift away from each other until eventually there is misalignment towards the side of the FOV.
\item Aberration: The aberration of the zoom lens increases towards the edge of its FOV and causes the focal spots to deviate from the diffraction-limited spot size. This causes the expansion of the OLED pixel image on the SLM plane, which couples part of the optical energy emitted from each OLED pixel into the surrounding SLM pixels that have incorrect transmissions for weight encoding. 
\end{enumerate}
The non-uniform transmission caused by the vignette could be fixed by making a pixel-wise LUT of the OLED display, which is described in Section \ref{vignette_correction}.
While there is no easy solution to nonlinear image distortion and aberration, we characterize them in Section \ref{image_quality}. 

\section{Correction of Optical Vignette}\label{vignette_correction}
To enable computation of large dot products, we corrected for intensity fall-off towards the edge of the FOV, caused by optical system vignettes as discussed in Section \ref{alignment}. The correction was especially important for optical fan-in (discussed in Section \ref{optical_fan_in}), since each pixel, regardless of its position in the FOV, should contribute the same amount of optical energy to the detector, if set at the same pixel value.

\begin{figure}[h!]
\includegraphics [width=0.8\textwidth] {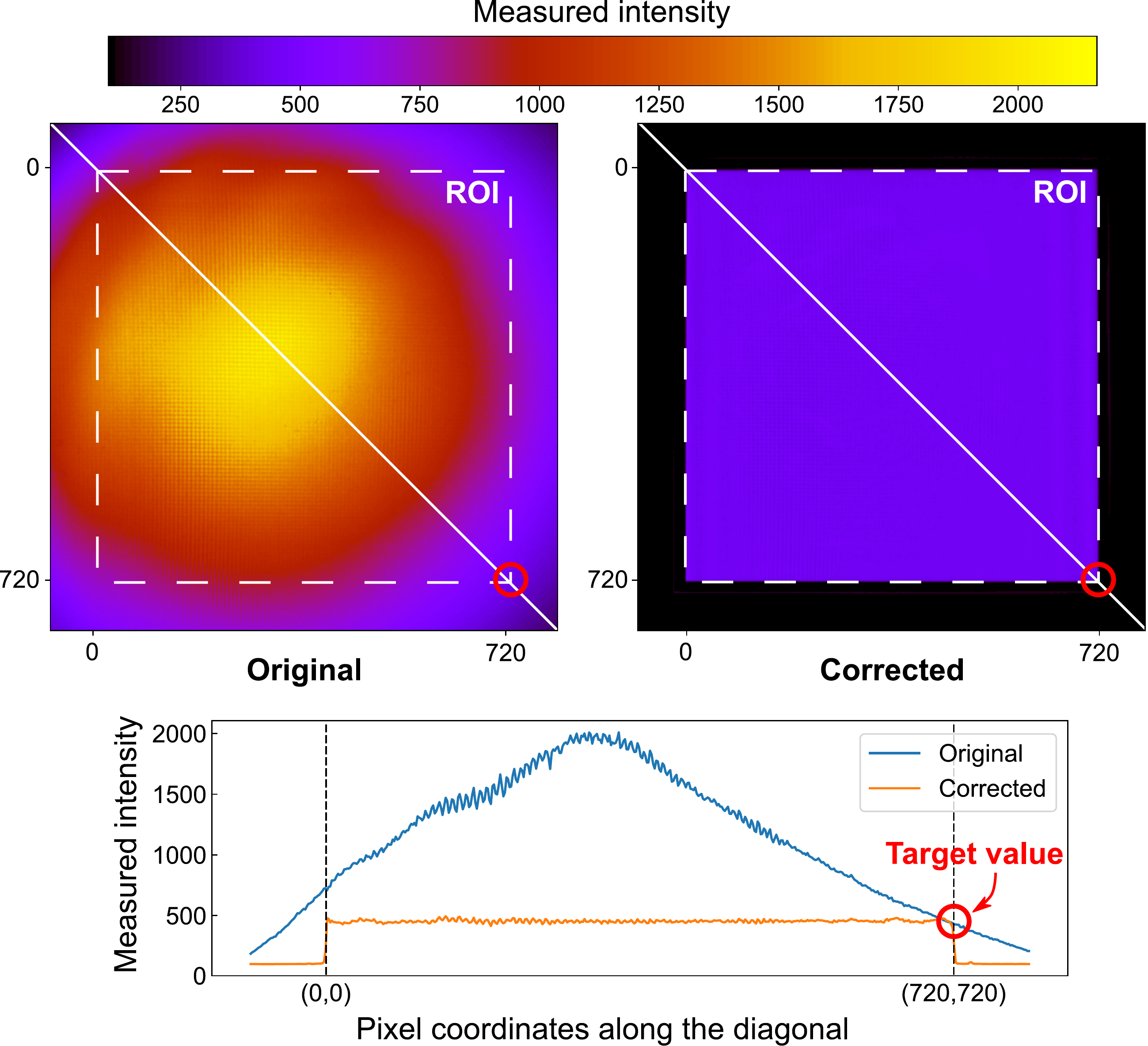}
\caption{\textbf{Correction of Non-uniform Transmission of the Optical System.} The original intensity distribution is shown in the top left panel (``Original"), with intensity falling off towards the edges of the region of interest (ROI). The correction procedure reduced the intensity of pixels near the center to match the target value near the darkest corner (720, 720) in the selected ROI. The image after correction is shown in the top right panel (``Corrected"). The bottom panel shows intensity along the diagonal pixels of the ROI (solid white lines) before and after correction.
}
\label{correction}
\end{figure}

Correction was performed by making an attenuation map that compensates for different transmissions of pixels at different locations. We first configured the OLED to display at the maximum pixel brightness uniformly across the entire FOV, and then captured an image of the OLED display at the detector plane. Due to the vignette effect of the optical system, the intensity distribution was not uniform at the detector plane (Fig. \ref{correction}, top left panel). A region of interest (ROI) was then chosen, in which we sought to achieve uniform intensity for all OLED pixels. The minimum intensity in the ROI was set as the target intensity value (circled areas in the top left panel of Fig. \ref{correction}). The brightness of other OLED pixels was reduced iteratively until the intensity of their images matched that of the target value. The result of the correction is shown in Fig. \ref{correction}, top right panel, where uniform intensity was achieved in an ROI of size $720 \times 720$. Meanwhile, an attenuation map was established to determine the percentage by which each OLED pixel should be attenuated in order to achieve a uniform output intensity.

\section{Pixel Walk-off and Crosstalk due to Imaging Imperfections} \label{image_quality}

\begin{figure}[h!]
\includegraphics [width=\textwidth] {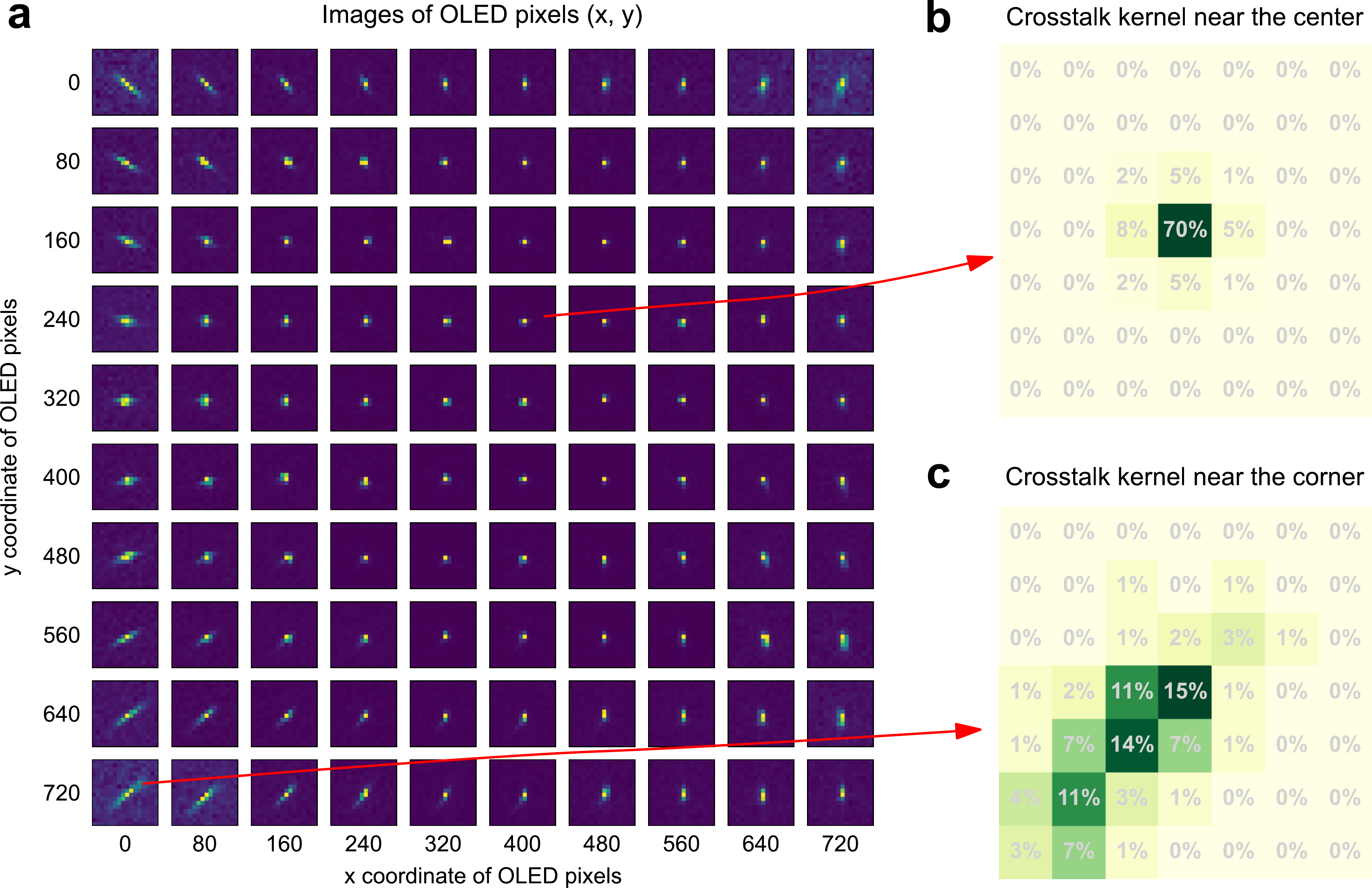}
\caption{\textbf{Imaging Quality of the OLED Pixels.} \textbf{a,} Images of the OLED pixels in a region of interest (ROI) of size $720 \times 720$. Only the pixels on the grid points of a 2D grid (with edge length of 80 pixels) were switched on. The OLED display was first imaged onto the SLM, and then relayed to the detector plane where the images were captured by a camera. The intensity inside each block was normalized to the maximum pixel value in the block. \textbf{b(c),} The $7 \times 7$ crosstalk kernels near the center (corner) of the FOV. Each entry denotes the percentage of optical power coupled into each individual SLM pixel, with all the power supposed to couple to the central pixel.}
\label{img_quality}
\end{figure}

We examined two aspects of the optical system's imaging quality: walk-off of pixel alignment due to image distortion, and degradation of focal spots due to aberration (both discussed in Section \ref{alignment}). To visualize these effects, we captured an image of a sparse 2D grid of pixels displayed on the OLED screen. The grid was composed of blocks with an edge length of 80 pixels, with only the pixel at the center of each block was turned on. Fig. \ref{img_quality}a shows how the imaging quality of single pixels changed across an ROI of $720 \times 720$ pixels. For example, images of a single pixel tended to be sharp and focused near the center of the FOV, while images of pixels towards the corner of the FOV spread out into a streak along the radial direction. This is probably due to coma aberration, which is common to most imaging systems (Fig. \ref{img_quality}a). Meanwhile, the walk-off of pixel alignment could be observed by the deviation of the focus from the center of each block (Fig. \ref{img_quality}a). As discussed in Section \ref{alignment}, pixel walk-off due to linear image distortion can be corrected with careful mechanical alignment, while nonlinear distortion cannot be eliminated. Based on Fig. \ref{img_quality}a, the pixel walk-off was insignificant for the ROI of $720 \times 720$. Incidentally, the largest possible ROI for optical matrix-vector multiplication was determined by the trade-off between optical power transmission and imaging quality distribution (Fig. \ref{correction} and Fig. \ref{img_quality}a). Since the imaging quality was better on the right side, the ROI was shifted slightly to the right of the brightest part of the intensity distribution.

Pixel walk-off and crosstalk both resulted in errors in weight modulation, due to the coupling of optical energy into neighboring SLM pixels with incorrect modulation weights. These effects were quantified and modeled as so-called crosstalk kernels, which are similar to convolution kernels but vary gradually in space. Fig. \ref{img_quality}b, c show the intensity distribution of a focal spot near the center (corner) of the FOV in a $7 \times 7$ block of SLM pixels, with the central pixel denoting the SLM pixel that the bright OLED pixel should align to. When both input vectors and weights were natural images (whose pixel value variation was smoother, and usually constitute the first layer of neural networks for image classification), the error caused by walk-off and crosstalk was less severe. For applications in optical neural networks (ONNs), such imaging errors were modeled during the training process by random affine transforms and 2D convolution to enhance the model's resilience to imaging errors (Section \ref{onn_training}).

\section{System Noise Characteristics} \label{noise_charac}

\begin{figure}[h!]
\includegraphics [width=0.6\textwidth] {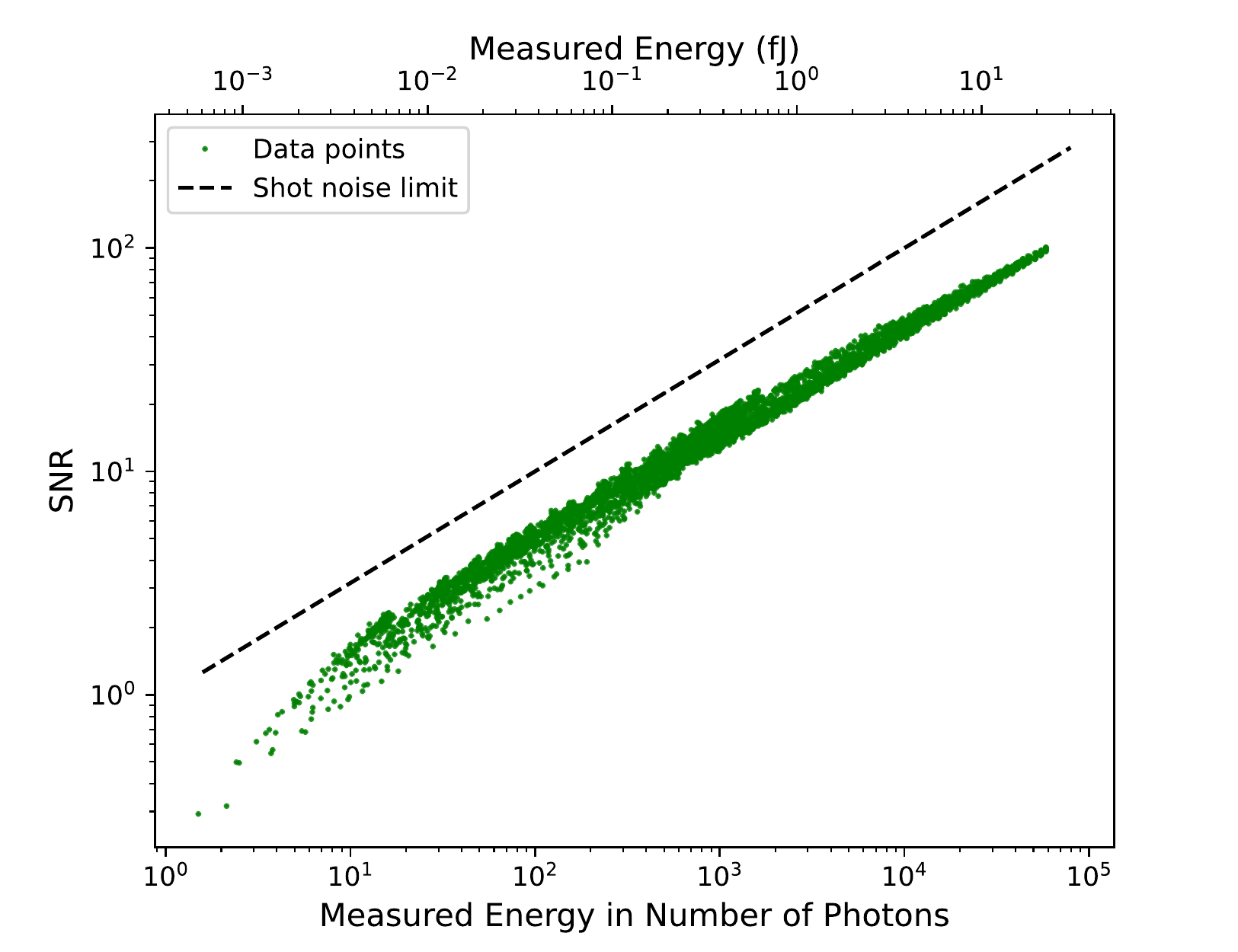}
\caption{\textbf{The Signal-to-noise Ratio (SNR) as a Function of Photon Flux, as Measured by the MPPC.} The integration window was \SI{150} {\nano\second} under analog mode. }
\label{SNR}
\end{figure}

We examined temporal fluctuations of each part of the system and describe hindrances in approaching shot noise-limited performance. Overall, when the OLED was set at a constant brightness and the SLM at a constant transmission across all pixels, the SNR of optical power measurements were about half the shot noise-limited SNR (Fig. \ref{SNR}). Sources of excess noise, in addition to shot noise, include intensity fluctuation of the OLED display, phase instability of the SLM, and the intrinsic noise of the detector. At high optical power, noise from external sources dominates the SNR measurement; as the power decreases, shot noise becomes a dominant source of noise. At extremely low optical power, the intrinsic noise of the detector is mainly responsible for deviations from the shot noise-limited performance.

\begin{figure}[h!]
\includegraphics [width=\textwidth] {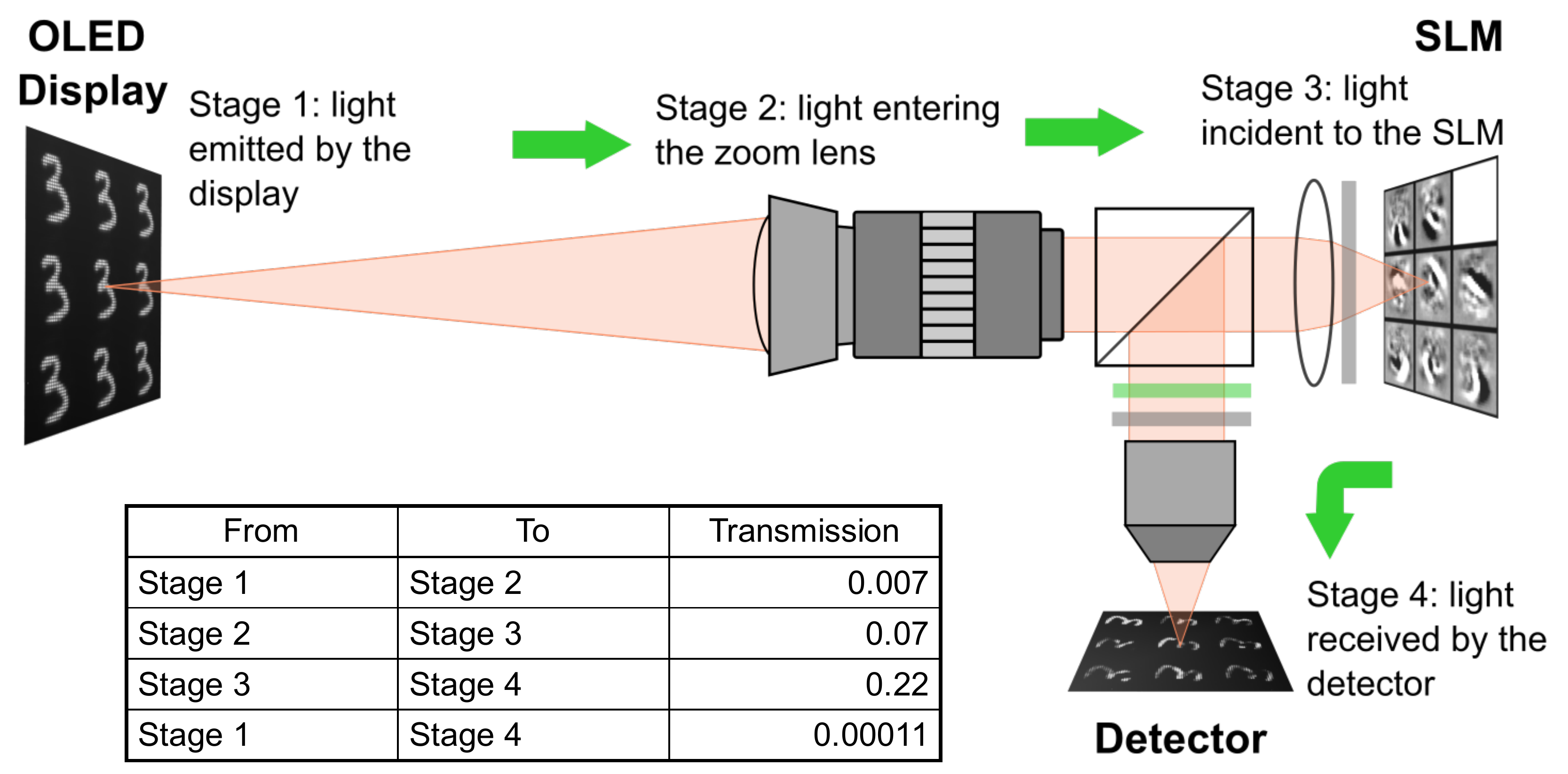}
\caption{\textbf{The Optical Power Transmission at Each Stage of the Setup.}  }
\label{setup_trans}
\end{figure}

There are three main components causing intensity fluctuations in OLED displays: raster scanning during screen refreshing, pulse width modulation for brightness control, and the thermal noise of OLEDs. When a stationary image was shown on the OLED display, no perceivable raster scanning pattern was observed in high-speed videos of the display. There were occasional black scanning stripes and short bursts of flashing, which are likely to have been caused by some refreshing mechanism. The OLED display did not seem to use pulse width modulation to adjust brightness until very low brightness settings were reached (below \SI{\sim 35}{\percent}). Therefore, we avoided setting pixels to low values, and instead used neutral density filters to attenuate light for extremely low-light measurements. The intensity fluctuation of the light source was mitigated by the high attenuation of the imaging system. Due to the large emission angle of OLED pixels ($>60$ degrees), most of the optical power was not collected by the zoom lens, which has a small collection angle. It was estimated that only \SI{\sim 0.7}{\percent} of light was collected by the zoom lens (Fig. \ref{setup_trans}). The high loss converted the thermal state of the OLED light closer to a coherent state by coupling vacuum states to it \cite{berchera2019quantum}, which improved the SNR and brought the ratio closer to the shot noise limit. It should be noted that even though the light collection efficiency was low for the zoom lens, the transmission from before the SLM to the detector (where the computation took place) was quite high at \SI{\sim 22}{\percent} (Fig. \ref{setup_trans}). Thus, optical energy efficiency can be drastically improved if the OLED and zoom lens are replaced with a stable coherent source and an optical system with high transmission in a customized setup.

The phase fluctuation of the SLM stems from the constant switching of voltage across the liquid crystal layers. This fluctuation could be measured by monitoring the intensity fluctuation of the laser diffraction pattern generated by a phase grating on the SLM. According to the manufacturer, the SLM used in this experiment oscillated at \SI{53} {\kilo\hertz}, with a measured peak-to-peak power ripple of 0.24\% for the first-order diffraction spot. Compared to OLED intensity fluctuations, instabilities caused by the SLM are relatively minor.

The intrinsic noise of the detector (e.g., thermal noise or dark counts) contributed to excess noise that was only apparent when the optical power to be measured was extremely weak. As discussed in Section \ref{MPPC}, the detector's intrinsic noise was negligible for high optical power ($\gg$\SI{1} {\pico\watt}). The analog noise floor becomes significant for low optical power at \SI{\sim 1} {\pico\watt}, which necessitates photon counting for even lower photon flux.

\section{Optical Fan-in and Detection Energy Consumption} \label{optical_fan_in}

The optical fan-in operation carries out the summation in vector-vector dot product computation by focusing optical spatial modes onto the active area of a detector. Unlike digital summation, which sums the readouts from different detector pixels with digital electronic circuits, optical fan-in implements summation through the accumulation of photoelectrons generated in the same piece of detector material. Optical fan-in reduces energy consumption by skipping the digital summation circuits and reducing the number of pixels together with associated amplification circuits. The energy consumption caused by charging and discharging of the detector can be reduced by shrinking the size as well as the capacitance of the detector through optical focusing. 

In this experiment, the optical fan-in was performed by an objective lens, which projected a de-magnified image of the SLM onto the active area of the MPPC. The de-magnification factor from the SLM to the detector was $0.276\times$, and the total de-magnification factor from the OLED display to the detector was $\sim 0.0442\times$. In other words, each OLED pixel of original pitch \SI{57.5}{\micro\metre} was imaged to a size of \SI{2.54} {\micro\metre} on the detector. Therefore, the entire $711 \times 711$ pixel array could be imaged onto the detector within a square size of $\SI{1.806} {\milli\metre} \times \SI{1.806} {\milli\metre}$, which fits into the $\SI{3} {\milli\metre} \times \SI{3} {\milli\metre}$ active area of the MPPC. During ideal optical fan-in, all spatial modes associated with a single dot product should be integrated by a single piece of detector material. In this experiment, since the MPPC consists of multiple photodiodes, not all the spatial modes were focused onto a single photodiode (however, it should be noted that the photocurrents of these pixels were still summed as analog signals to form a single output, and there was no digital operation involved). Each photodiode covered $(\SI{50} {\micro\metre}/\SI{2.54} {\micro\metre})^2 \sim 388$ spatial modes for in-pixel summation. Even with so many spatial modes, no saturation was observed. This is because the photon flux was extremely low for each spatial mode, and thus the simultaneous arrival of multiple photons was rare. The photoelectrons generated by different photodiodes were superimposed as a single analog voltage signal at the detector output.

The lower bound of energy consumption of optical fan-in is determined by the capacitive noise of the detector \cite{hamerly2019large}, and can be viewed as the energy cost of analog summation/accumulation. In this experiment, the MPPC worked at a typical wall-plug power of \SI{1.1}{W}. This came out to \SI{0.3} {\milli\watt} consumed for each photodiode, after dividing the wall-plug power by the total number of the photodiodes (3,600). If the data input and output rates were matched to the \SI{3} {\mega\hertz} bandwidth of the detector, each photodiode would consume $\SI{0.3} {\milli\watt} /  \SI{3}{\mega\hertz} = \SI{100} {\pico\joule}$ per update cycle. Dividing this by the number of spatial modes captured by each photodiode yields the energy consumed for each addition. This comes out to $\SI{100} {\pico\joule} / 388 = \SI{258} {\femto\joule}$.  

In principle, the energy budget for each addition can be reduced by using smaller focal spots, lower detector gain, and increasing the number of spatial modes for in-pixel summation. The focal spot can be reduced to the diffraction limit of $\lambda/2$ in air (Abbe's resolution limit with a numerical aperture of 1), which makes each spatial mode occupy $(\lambda/2)^2=(\SI{532} {\nano\metre}/2)^2=\SI{0.071} {\micro\metre\squared}$ area on the detector. The energy consumption for each addition scales with area, which would be $\SI{258} {\femto\joule} \times (\SI{0.532} {\micro\metre}/2/\SI{2.54} {\micro\metre})^2 = \SI{2.83} {\femto\joule}$. The area of each spatial mode can be further reduced by focusing light in materials with refractive index larger than 1 (e.g., glass). 

In this experiment, the photodiodes were operated with a high gain (Geiger mode) to provide extremely high SNR for single photon detection. In practice, a lower gain can be better for two reasons: 1. The energy cost per addition can be further reduced with a lower gain. 2. A lower gain prevents saturation of the detector, and allows the accumulation of more terms in a dot product, which is essential for an optical energy advantage. Ideally, the detector volume should be minimized to reduce capacitance, which in turn reduces thermal noise and potentially allows a high voltage across the detector to exempt the use of amplifiers \cite{hamerly2019large, miller2017attojoule}, Meanwhile, the number of carriers scales with the detector volume, which results in a limited full well capacity for a small detector volume. A high gain results in each photoelectron being amplified into many electrons and depletes the carriers in the detector volume quickly. For this reason, the gain of the detector should be reduced to be just enough to amplify the signal to overcome thermal noise. For a detector area of only one spatial mode (assuming $\lambda/2$ focal spot size at \SI{525} {\nano\metre}), the RMS value of thermal noise electrons is calculated as $\sqrt{kTC} = \sqrt{\SI{4.14e-21} {\joule} \times \SI{27.5} {\atto\farad}}= \SI{3.38e-19} {\coulomb} = 2.1 e^-$ at room temperature \cite{hamerly2019large, miller2017attojoule}. At the detection level of 1 photon per spatial mode, a moderately low gain can be applied such that the number of electrons generated by each photon is greater than the number of noise electrons.

Besides applying gains, the SNR of photon detection can also be improved by increasing the number of spatial modes for in-pixel summation. Even though both detector area and capacitance scale with the number of spatial modes to be summed ($N$), the overall noise scales with $\sqrt{N}$ while the total signal photons scale with $N$. For example, with 0.5 signal photons (without any detector gain) and 2 noise electrons per spatial mode, the $\text{SNR}=0.25$ for each spatial mode; with 10,000 spatial modes, the SNR can be enhanced to $0.25 \times \sqrt{10,000}=25$. Therefore, even in the thermal noise-limited (as opposed to shot noise-limited) case, it is possible to anticipate less than 1 photon consumed for each accumulation in the dot product computation, as long as the vector size is sufficiently large.

Based on the optimization measures mentioned above, the energy consumption for optical fan-in can be reduced to \SI{100} {\atto\joule} per addition, which is sufficiently low to enable optical processors at least $10^2$ times more energy-efficient than state-of-art machine learning accelerators (e.g., at \SI{\sim e2} {\femto\joule} per multiply-and-accumulate operation (MAC) = one multiplication plus one addition operation) \cite{reuthersurvey2020}. 

It is noted that optical fan-in not only implements the summation with physical accumulation of electrons, but also reduces the memory usage for storing intermediate results. For running typical neural networks, memory-associated energy costs often account for the majority of the total energy cost, higher than the energy spent on arithmetic operations \cite{sze2017efficient}. Even though digital electronic processors are improving in energy efficiency by tailoring data flow to machine-learning tasks \cite{sze2017efficient} and by incorporating more on-chip memory \cite{jouppi2017datacenter}, there is still a significant amount of memory that can be further reduced. For example, in systolic arrays, vector-vector dot products are computed by adding $w_ix_i$ $(i=1...N)$ term by term to a partial sum, which requires $N$ dedicated memory units each to read and write the partial sum once. Digital electronic adders usually have a small number of input operands, and cannot perform the summation of a large number of terms without saving the intermediate results. In comparison, optical fan-in can implement the summation of a large number of terms ($10^3 \text{-} 10^5$) physically in a single step, which exempts the use of any intermediate memory units, potentially leading to a substantial amount of energy saved.   

\section{Comparison to the Stanford Matrix-vector Multiplier} \label{1d_vs_2d_svm}

\begin{figure}[h!]
\includegraphics [width=0.8\textwidth] {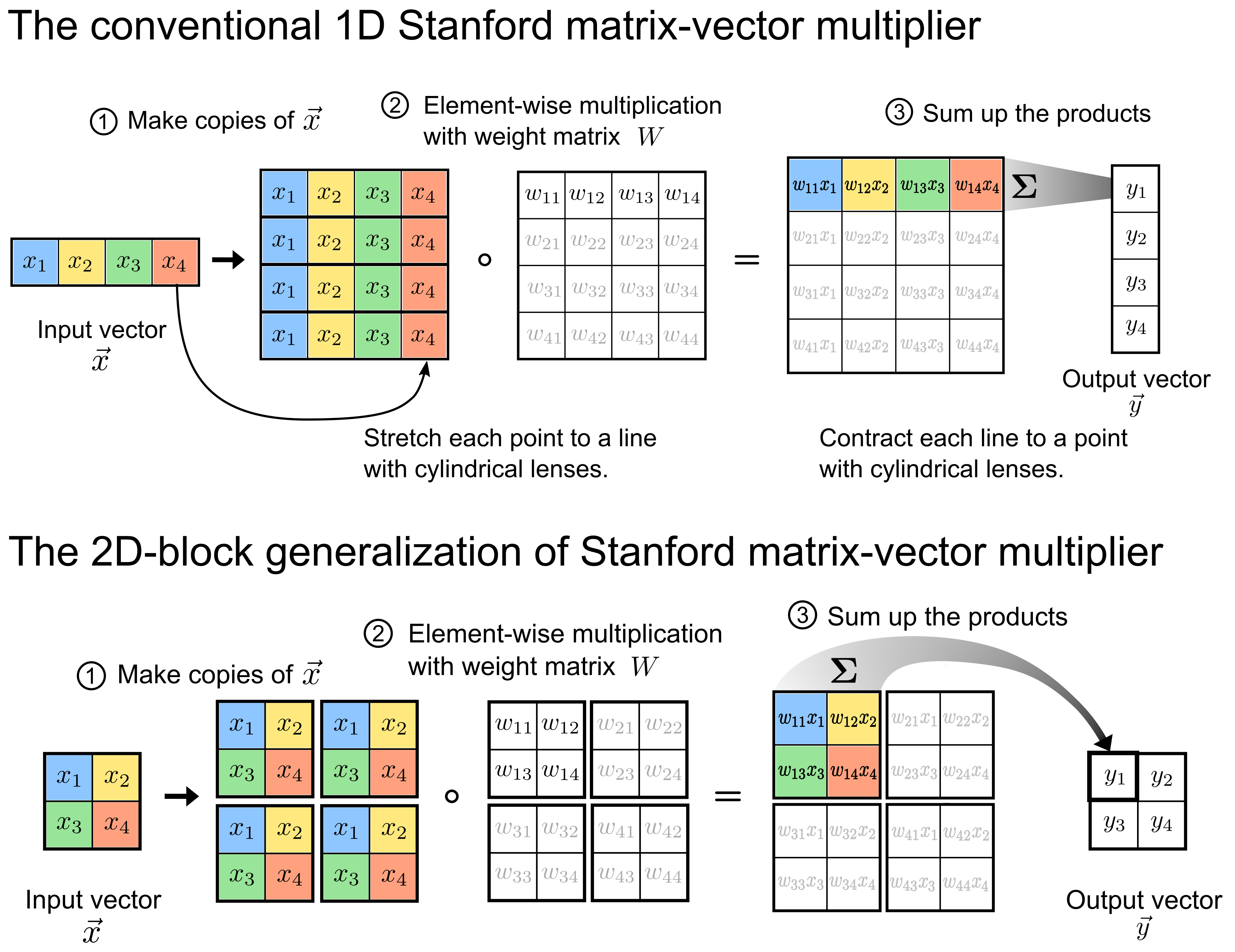}
\caption{\textbf{The Comparison between the Classical Stanford Matrix-vector Multiplier and our 2D-block Scheme.} }
\label{fig_1d_vs_2d}
\end{figure}

Our setup can be viewed as a generalization of the classical Stanford matrix-vector multiplier, whose operation can be summarized in three steps: fan-out, element-wise multiplication, and fan-in (Fig. \ref{fig_1d_vs_2d}). The major difference between our setup and the Stanford matrix-vector multiplier lies in the geometric arrangement of the input vector and its corresponding weights (the weight vector) in 2D blocks instead of 1D arrays, which lead to different physical implementations. Compared to 1D arrays, the 2D-block arrangement of both input and weight vectors offer advantages in scalability, especially for vector-vector dot products with a large vector size.

\begin{figure}[h!]
\includegraphics [width=0.9\textwidth] {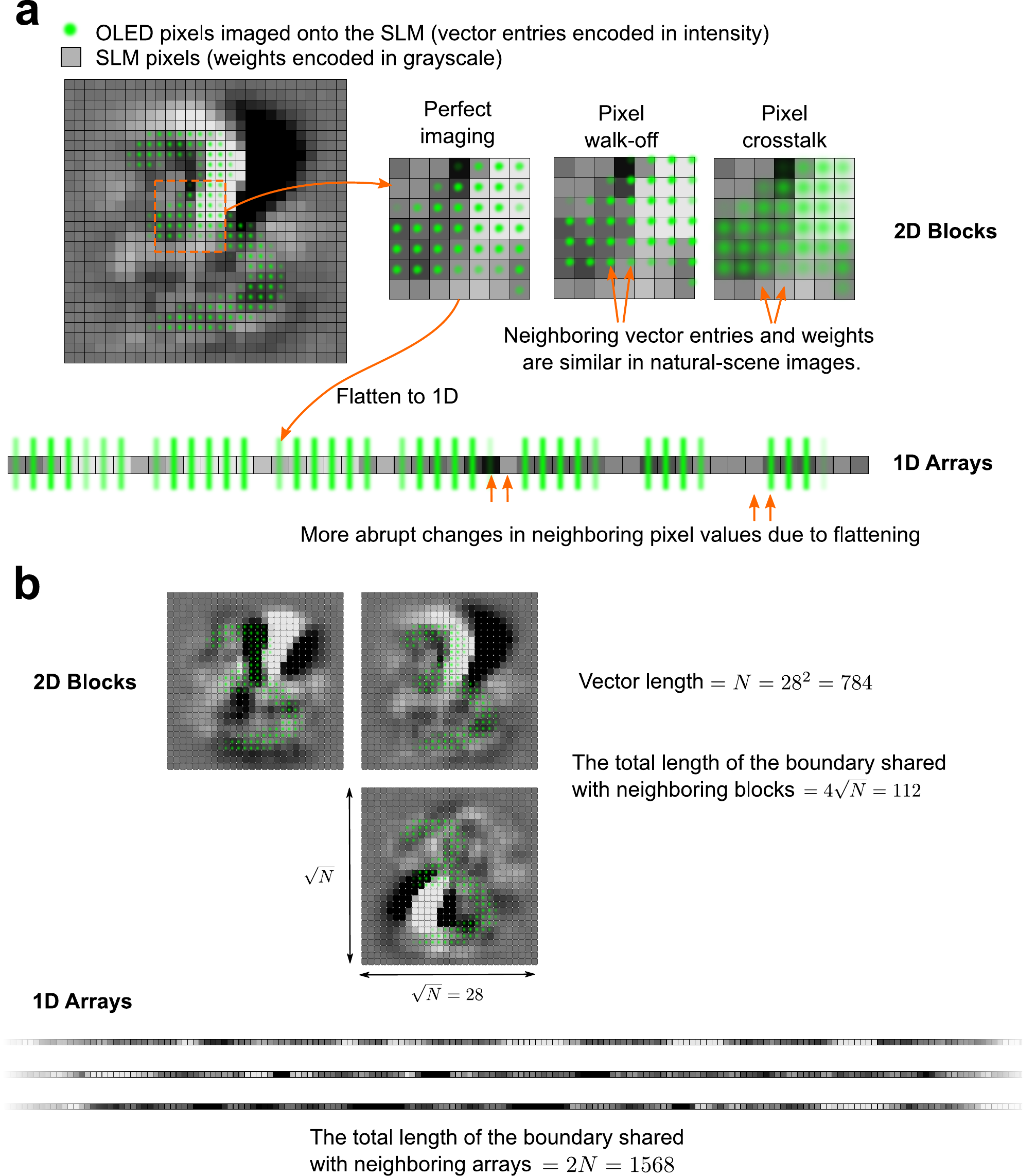}
\caption{\textbf{Comparison between the 2D-block and 1D-array Representation of the Input Vector and Weight Matrices.} \textbf{a,} Error reduction effects of 2D-block representation of natural-scene images. The smooth local features intrinsic to the image data and corresponding weight matrices mitigate the crosstalk between neighboring pixels caused by imperfections in imaging (e.g., degradation of focal qualities) or instrumentation (e.g., crosstalk between SLM pixels). Reshaping 2D images into 1D arrays breaks the continuity intrinsic to these images. \textbf{b,} The 2D-block representation reduces the crosstalk between neighboring vector and weight blocks. The 2D blocks share substantially shorter boundaries with each other ($\propto \sqrt{N}$) compared to 1D arrays ($\propto N$), which helps to reduce crosstalk between different vector and weight blocks, or to reduce the gaps between them for a higher fill factor on the SLM. }
\label{fig_errors_1D_vs_2D}
\end{figure}

The 2D-block arrangement is especially suitable for applications in image classification, where the input data are usually 2D images of high dimensions. For natural-scene images, and the weights trained for them by neural networks, (e.g., see Fig. \ref{digit_mod}), the pixel values usually do not vary drastically in local areas except for few high-contrast boundaries. Preserving smooth local features in the 2D form reduces errors resulting from minor shifting or blurring of the input and weight vectors. Therefore, our setup is more tolerant of minor errors in instrumentation, such as imperfect imaging or crosstalk between SLM pixels (Fig. \ref{fig_errors_1D_vs_2D}a). In comparison, flattening the vectors and weights into 1D arrays, as required by the Stanford matrix-vector multiplier, would break the smoothness of local features, which can result in more severe errors (e.g., the crosstalk level between the neighboring SLM pixels increases with the level of contrast between them). 

The 2D-block arrangement also reduces the amount of crosstalk between different input or weight vectors tiled next to each other (Fig. \ref{fig_errors_1D_vs_2D}b). For each block, crosstalk occurs to the pixels in contact with adjacent blocks. When a vector of size $N$ is wrapped into a square, the total number of pixels on its perimeter equals $4\sqrt{N}$ (Fig. \ref{fig_errors_1D_vs_2D}b top panel). If the vector were instead shaped into a 1D array, it would share a boundary of a total of $2N$ pixels with neighboring arrays (Fig. \ref{fig_errors_1D_vs_2D}b bottom panel). For a large vector of size $N$, 2D blocks can potentially reduce the amount of crosstalk by over an order of magnitude more than 1D arrays, due to the difference in scaling of $\sqrt{N}$ (Fig. \ref{fig_errors_1D_vs_2D}b). Even though a gap can be introduced between the regions corresponding to different vectors, it would also reduce the fill factor as well as the computational throughput. Therefore, through decreasing the surface-to-volume ratio, the 2D-block arrangement can effectively reduce crosstalk between different vectors with a high fill factor on both OLED and SLM panels. 

It should be noted that in the classical Stanford matrix-vector multiplier both optical fan-out and fan-in are implemented with cylindrical lenses as reverse processes of each other. In the 2D-block arrangement, this symmetry is broken because the optical fan-out process needs to create copies of the same block, while the optical fan-in focuses all elements in each block. In our demonstration, we performed the fan-out operation digitally by displaying different copies of the same vector on the OLED display, which was sufficient for the purpose of demonstrating sub-photon scalar multiplication. Purely optical fan-out is necessary to achieve the total energy benefit for optical matrix-vector multiplication (for details, see Section \ref{scaling}). For the 2D-block scheme, it is possible to implement optical fan-out with several techniques, including imaging with a microlens array \cite{andregg2018wavelength, hayasaki1992optical} or beam splitting using an array of beam splitters \cite{hemmi2013development}. The optical fan-out operation would allow a very low number of photons in each spatial mode: the optical fan-out splits the photons emitted from a reasonably weak light source and distribute each portion of the photons into many spatial modes, with each mode performs a scalar multiplication. In addition, since each spatial mode is the result of many vacuum modes being coupled to the original spatial mode populated by a single light source, the noise of each spatial mode can be reduced to be quite close to shot noise, even if the light source intensity is not perfectly stable.  

\clearpage
\part{Vector-Vector Dot Product Precision}

We characterized the precision of our optical matrix-vector multiplication by computing vector-vector dot products, which constitute general matrix-vector multiplication. The answer $y$ (scalar) to the dot product between \textit{input vector} $\vec{x}$ and \textit{weight vector} $\vec{w}$ is defined as:

\begin{equation} \label{eq:dot_product1}
    y=\vec{w}\cdot\vec{x}=\sum_{k=1}^{N} w_k x_k.
\end{equation}
where $x_k$, $w_k$ are in general real numbers.

\section{Computing Dot Products with Signed Elements using Incoherent Light} \label{mvm_conversion}

Our setup can only perform dot products between vectors with non-negative elements, because $x_k$ is encoded with the intensity of each spatial mode, and $w_k$ is encoded with the transmission of each spatial mode. However, dot products between vectors of signed elements can always be reduced to those between non-negative-valued vectors with minimal digital processing overhead \cite{goodman1978fully}. For a general vector with signed elements $\vec{x}^{\text{signed}}$, where each element $x_k^{\text{signed}} \in [x_{\text{min}}^{\text{signed}}, x_{\text{max}}^{\text{signed}}]$, $x_{\text{min}}^{\text{signed}}, x_{\text{max}}^{\text{signed}} \in \mathbb{R}$, a non-negative vector $\vec{x}'$ can be obtained by adding bias terms and rescaling:

\begin{equation}\label{eq:dot_product2}
    \vec{x}' = \frac{x_{\text{max}}'-x_{\text{min}}'}{x_{\text{max}}^{\text{signed}}- x_{\text{min}}^{\text{signed}}} \; \vec{x}^{\text{signed}} + \frac{x_{\text{max}}^{\text{signed}}x_{\text{min}}'-x_{\text{min}}^{\text{signed}}x_{\text{max}}'}{x_{\text{max}}^{\text{signed}} - x_{\text{min}}^{\text{signed}}} \; \vec{1},
\end{equation}
where $\vec{1}=(1,1,\cdots,1)$ is a constant vector of size $N$. It can be verified that $x_k'$ is tightly bounded between $[x_{\text{min}}', x_{\text{max}}']$ with $x_{\text{max}}' \geq x_{\text{min}}' \geq 0$. Therefore, two vectors with signed elements $\vec{w}^{\text{signed}}$ and $\vec{x}^{\text{signed}}$ can be converted to non-negative vectors $\vec{w}'$ and $\vec{x}'$, and the dot product $\vec{w}^{\text{signed}}  \cdot \vec{x}^{\text{signed}}$ equals to the linear combination of $\vec{w}' \cdot \vec{x}'$, $\vec{w}' \cdot \vec{1}$, $\vec{1} \cdot \vec{x}'$ and a constant term. All three dot products are between vectors of non-negative elements. The dot product between $\vec{1}$ and any vector equals the summation of all of the vector elements, which were computed optically like any other dot product. 

In machine learning applications, the input vector $\vec{x}'$ is either the input to the neural network or the neural activation of the previous layer, both of which are non-negative values after the ReLU nonlinear activation function. Therefore, since $\vec{x}^{\text{signed}}$ is already non-negative with $x_{\text{min}}^{\text{signed}}=0$,

\begin{equation}\label{eq:dot_product_sum}
    \vec{w}^{\:\text{signed}} \cdot \vec{x}^{\:\text{signed}} = c_1 \ \vec{w}' \cdot \vec{x}' + 
    c_2 \ \vec{1} \cdot \vec{x}'.
\end{equation} 

 For simplicity, both $\vec{x}'$ and $\vec{w}'$ can be normalized to the range $[0,1]$, and the coefficients in Eq. \ref{eq:dot_product_sum} can be solved as: $c_1=(w_{\text{max}}^{\text{signed}}-w_{\text{min}}^{\text{signed}})x_{\text{max}}^{\text{signed}}$, $c_2=w_{\text{min}}^{\text{signed}} x_{\text{max}}^{\text{signed}}$. The normalized vectors $\vec{x}'$ and $\vec{w}'$ were loaded onto the OLED display and the SLM, respectively, according to the hardware LUTs (e.g., Fig. \ref{OLED}c and Fig. \ref{amp_mod}b). In reality, the SLM could not achieve zero transmission which led to a minimum modulation $x_{\text{min}}^{\text{signed}}=\epsilon=0.02$ (Fig. \ref{amp_mod}b). In other words, $w_k'$ was normalized to the range $[\epsilon,1]$ instead of $[0,1]$. In this case, $c_1=\frac{1}{1-\epsilon} (w_{\text{max}}^{\text{signed}}-w_{\text{min}}^{\text{signed}})x_{\text{max}}^{\text{signed}}$, $c_2=\frac{1}{1-\epsilon} (w_{\text{min}}^{\text{signed}}-\epsilon w_{\text{max}}^{\text{signed}})x_{\text{max}}^{\text{signed}}$. 
 
 In summary, the dot product between two vectors with signed elements can be converted to two dot products between non-negative vectors, which can be first computed purely with optics, and then combined with only 2 digital multiplications and 2 digital additions. In other words, the price of conversion is a doubling of the amount of optical computation with a constant digital overhead, independent of vector size $N$. 

For matrix-vector multiplication, the computational overhead can be further reduced, since $\vec{1} \cdot \vec{x}'$ remains the same for the dot product between $\vec{x}$ and any row vector of the matrix. As a result, $\vec{1} \cdot \vec{x}'$ only needs to be computed once optically and can be reused afterwards. In other words, to compute a matrix of size $N' \times N$ multiplied with a vector of size $N$, in addition to the $N'$ optical dot products (which constitute $NN'$ MACs), only one additional optical dot product ($\vec{1} \cdot \vec{x}'$) is required. Thus, the amount of digital overhead is on the order of $O(N')$.

\section{Characterization of Dot Product Accuracy with Varying Photon Budget} \label{dprod_calibration}

\subsection{Generation of Test Datasets} \label{gen_rand_imgs} To generate a test dataset representative of general dot products, we randomly generated vector pairs $\vec{x}$ and $\vec{w}$ based on natural scene images from the STL10 dataset. Each vector was generated from a single color channel of one or more images patched together, depending on the target vector size (each image of size $L \times L$ contributes $N=L^2$ elements to the vector). We chose natural images since they are more representative of the inputs in image classification with globally inhomogeneous and locally smooth features. To adjust the sparsity of the vectors, different thresholds were applied to the image pixel values such that the dot product results cover a wider range of possible values. This was achieved by shifting the original pixel values (float point numbers normalized to the range $0\text{-}1$) in the entire image up or down by a certain amount, unless the value was already saturated at 1 (the maximum) or 0 (dark). For example, a shift of -1 would make the whole image dark. A shift of +0.2 would make all the pixel values that were originally larger than 0.8 saturated, and would increase all other pixel values by 0.2. This method allowed us to tune the overall intensity of the modulated images without losing the randomness of the distribution.

\begin{figure}[h!]
\includegraphics [width=\textwidth] {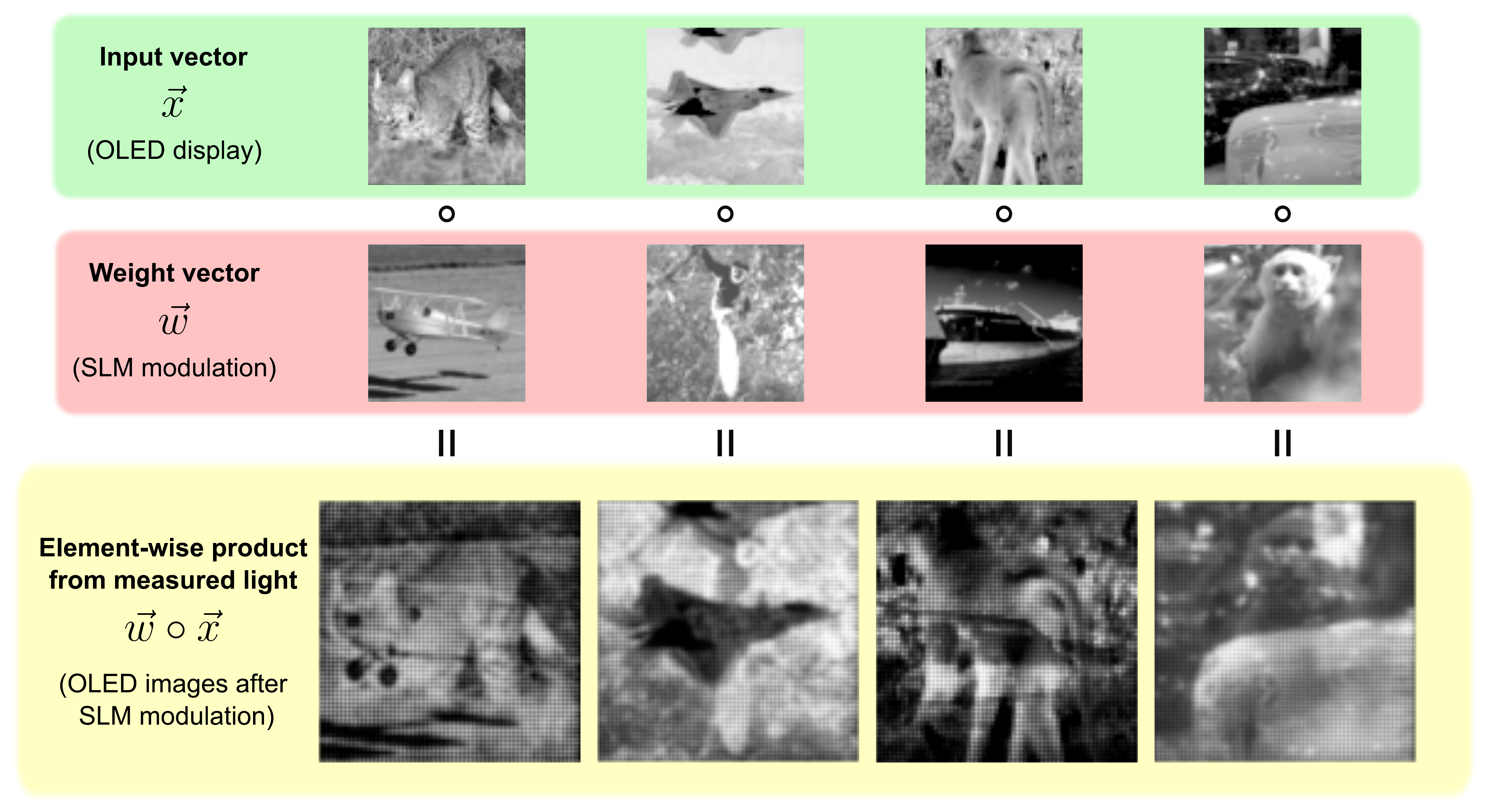}
\caption{\textbf{Example Measurement of Element-wise Multiplication between Random Vectors.} The top two rows show the corresponding input vectors on the OLED display and the weight vectors on the SLM. The bottom row displays modulated light captured by a camera. The input and weight vectors were generated from images in the STL10 dataset of size $64 \times 64$ (or $64 \times 64=4,096$ elements). Individual pixels are visible in the captured images.
}
\label{stl_images}
\end{figure}

The computation of dot products on the setup followed the same steps of element-wise multiplication and optical fan-in, as described in the main text. Fig. \ref{stl_images} shows a few more examples of element-wise multiplication, similar to Figure 2a in the main text.

 \subsection{Data Collection Scheme and Photon Budget Control} In order to study how dot product accuracy changes with photon budget, we used a sensitive detector (MPPC) to measure the integrated optical energy. The optical energy consumed for each dot product computation was controlled by tuning the detector integration times (e.g., Fig. \ref{SNR} had an integration time of \SI{150} {\nano\second}). To get enough statistics for noise distribution under low optical power, each detector readout measurement was repeated $T$ times for each vector pair $\vec{w}$ and $\vec{x}$. To get error statistics representative of general vector pairs, we also repeated the measurement for $S$ randomly generated vector pairs of different sparsity from randomly chosen images, as discussed in Section \ref{gen_rand_imgs} and Fig. \ref{stl_images}). We call this set of vector pairs the calibration dataset, and collected a total of $S \times T$ data points. Detector readout $v_{i,j}$ denotes the $j$th ($j = 1,2,\dots,T$) measurement made on the $i$th ($i = 1,2,\dots,S$) vector pair. For each vector pair, the mean value of the detector readouts $\bar{v}_i=1/T \sum_{j=1}^{T} v_{i,j}$ was calculated for large enough $T$ to eliminate the noise. The detector readouts were quantified either in optical energy, or in number of photons, which is optical energy divided by the photon energy (i.e., \SI{\sim 0.4} {\atto\joule} at \SI{525} {\nano\metre}). To enable energy efficiency comparisons between different vector sizes, the total optical energy, or number of photons, detected for each dot product was further divided by the number of multiplications in the dot product.  
 
 \begin{figure}[h!]
\includegraphics [width=0.6\textwidth] {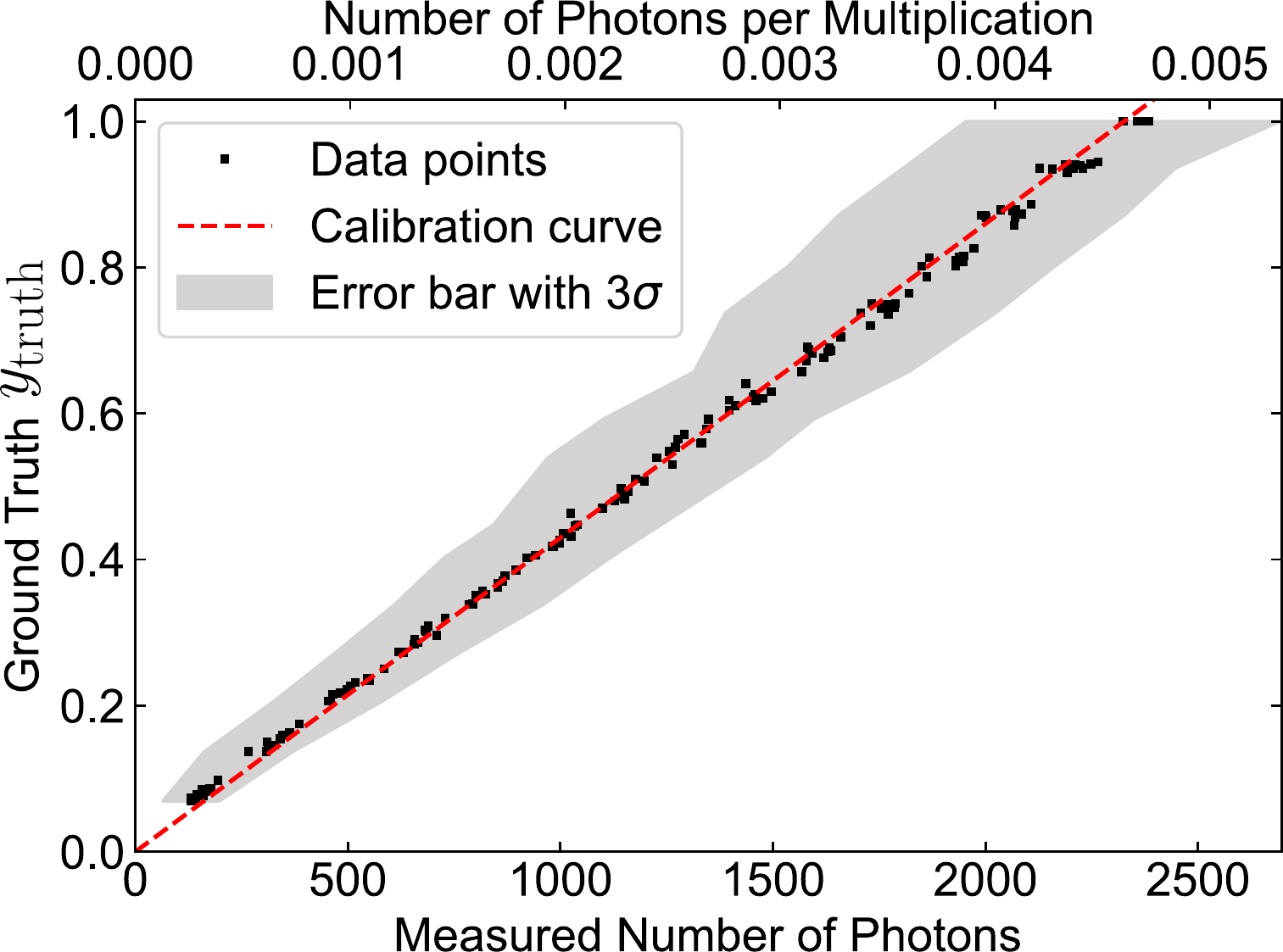}
\caption{\textbf{A Calibration Curve Converting Detector Readouts to Dot Product Results.} The mean detector readout ($\bar{v_i}$, x axis) is plotted against the corresponding ground truth ($y_{\text{truth}, i}$, y axis) for every vector pair in the calibration dataset. The unit of detector readout is either the number of photons (bottom axis) or the number of photons per multiplication (top axis). The vector length was $711 \times 711$ ($N=505,521$). The calibration curve (red dashed line) was obtained using a least-squares fit to the data points. The shaded area indicates 3 standard deviations of the repeated measurements. The average number of photons per multiplication of the data points in the plot is 0.0025.
}
\label{cali_curve}
\end{figure}

\subsection{Calibration of Detector Readouts} A calibration model $f$ was made to convert the average detector readout $\bar{v}_i$ to the dot product result $ y_{\text{meas},i}$ as $y_{\text{meas}, i} = f(\bar{v}_i)$. The calibration involved plotting the ground truth of the dot product $y_{\text{truth}, i}= \vec{x}_i\cdot \vec{w}_i$ versus $\bar{v}_i$, followed by fitting the data points to a linear curve using a least-squares criterion. Fig. \ref{cali_curve} shows an example of data points measured on vector pairs of length $N=505521$. The calibration curve $f$ is plotted in the dashed red line. The range of $y_{\text{truth}}$ was normalized to $[0,1]$ by rescaling $\vec{x}'$ and $\vec{w}'$, based on their definitions in Eq. \ref{eq:dot_product2}, with a multiplicative factor $1/\sqrt{N}$. With the calibration model, we could read out the dot product result based on the detector readout value. In principle, the calibration only needed to be performed once with the calibration dataset, unless the setup changed (e.g., adding extra attenuation) or has drifted over time.  

\subsection{Quantification of Single-Shot Dot Product Computation Error}

\begin{figure}[h!]
\includegraphics [width=\textwidth] {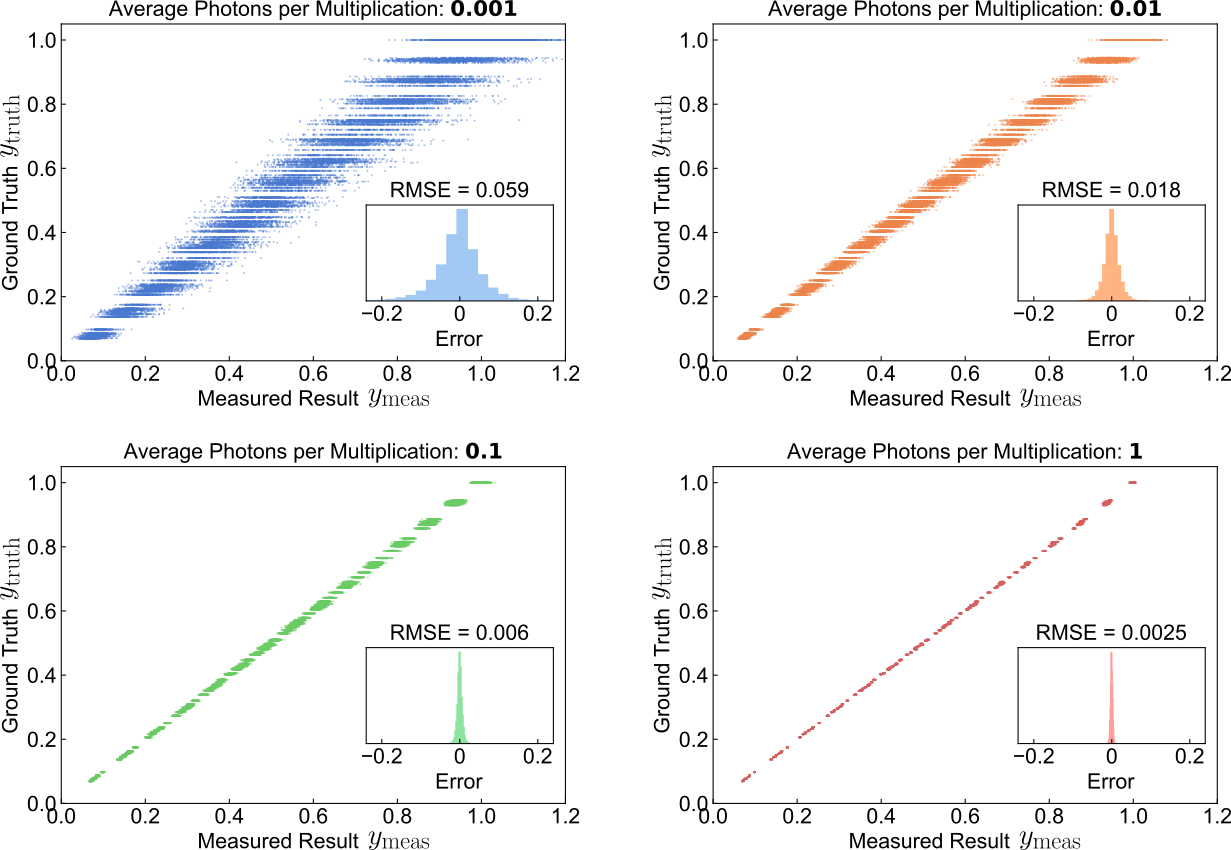}
\caption{\textbf{Dot Product Error Analysis at 4 Typical Photon Budgets.} The dot products were computed with vector size $N=711 \times 711=505,521$). The average number of photons per multiplication (indicated at the top of each plot) was controlled by the integration time of the MPPC detector, and averaged over the entire test dataset. For each vector pair, the measurement $y_{\text{meas},i,j}$ was repeated multiple times, and all the data points were plotted. The ground truth $y_{\text{truth}, i}$ is plotted against the corresponding measurements $y_{\text{meas},i,j}$. The histogram of errors $y_{\text{truth}, i}-y_{\text{meas},i,j}$ is shown in each inset. The overall accuracy representative of the dataset is characterized by the root-mean-square error (RMSE), which is also similar to the value of the standard deviation of the error distribution. The color code is the same as that in Figure 2 in the main text.
}
\label{error_hist}
\end{figure}

After obtaining the calibration curve, we generated another random vector pair test dataset in order to quantify the error statistics of dot product computation performed by our setup. Error was defined as the difference between the measured result and ground truth $y_{\text{truth}}-y_{\text{meas}}$. Suppose we have $S$ vector pairs in the data set and each is repeated $T$ times. For each detector readout $v_{i,j}$ ($i = 1,2,\dots,S$, $j = 1,2,\dots,T$), $\text{Error}_{i,j} = y_{\text{truth}, i}-y_{\text{meas},i,j}$, where $y_{\text{meas},i,j}=f(v_{i,j})$. Unlike calibration, here we used single-shot readouts $v_{i,j}$ rather than the mean value $\bar{v}_i$.

For each vector pair $\vec{x}_i$ and $\vec{w}_i$, the root-mean-square (RMS) error for different measurement trials was calculated as $\text{RMSE}_i=\frac{1}{T} \sqrt{\sum_{j=1}^{T} \text{Error}_{i,j}^2}$, which can be interpreted roughly as the most likely error one would get from a single-shot computation for vector pair index $i$. The total RMS error across different vectors in the dataset was calculated as $\text{RMSE}=\frac{1}{S} \sqrt{\sum_{i=1}^{S} \text{RMSE}_{i}^2}=\frac{1}{ST} \sqrt{\sum_{i,j} \text{Error}_{i,j}^2}$, which could be interpreted as the most likely error one would get from a single-shot computation by randomly selecting a vector pair from the entire test dataset. Histograms of the errors of the test dataset are shown in Fig. \ref{error_hist} insets. 

The scatter plots in Fig. \ref{error_hist} show how well the computed dot product results matched the ground truth, under different photon budgets. The average number of photons detected during these experiments were different, since they were determined by detector integration time. With higher photon budgets, the error decreases as the noise contribution to the error decreases. For a higher photon budget (\textgreater1 photon per multiplication), the RMS error stops decreasing and is instead limited by systematic error due to imperfections in the setup. The four scatter plots in Fig. \ref{error_hist}) correspond to the total RMSE data point of the same color in Figure 2b in the main text.
\clearpage

\part{Optical Neural Network for Image Classification}

\section{Training Protocol of Noise Resilient Optical Neural Networks} \label{onn_training}

For handwritten digit classification (MNIST database), we trained a 3-layer neural network with full connections (i.e., a multilayer perceptron, MLP). The inputs are 8-bit grayscale images of size $28 \times 28 = 784$ total pixels, followed by two fully-connected hidden layers, each comprising 100 neurons with ReLU as the nonlinear activation function. The output layer has 10 neurons, with each neuron corresponding to a digit from 0 to 9. The neural network was implemented and trained in PyTorch (1.7.0). To improve the robustness of the model to numerical error, we employed several techniques during training: 

\begin{enumerate}
    \item Quantization-aware training (QAT): The activation of neurons were quantized to 4 bits and weights to 5 bits to adapt to the numerical precision of the setup. For example, for vector size of 784, we found the SNR of dot products is equivalent to $\sim$4-bit numerical precision. Even though the SLM could be controlled with 8-bit numbers, we decided to quantize its weight to 5 bits, which matched better with its extinction ratio of 50 and did not seem to have any negative impact on MLP accuracy. To reduce the numerical sensitivity caused by quantization (e.g., in the regular nearest neighbor scheme, 0.49 is rounded down to 0, while 0.51 is rounded up to 1), we used random digitization between the adjacent levels, which was observed to improve the model robustness against random noise. We found that a few (6-12) warm-up training epochs with full 32-bit float precision helped to protect the model from aggressive quantization in the initial stage, after which the application of quantization noise was less likely to derail the training process, but still helped to fine-tune the parameters.
    \item Data augmentation with random image transform and convolution: To improve model tolerance to potential hardware imperfections, we imitated similar kinds of errors on input images. For example, the misalignment was modeled as random rotation (within $\pm 5^{\circ}$) and translation ($\pm 4\%$ of image size in any direction), mismatched zoom factor as random zooming ($\pm 4 \%$ of image size), and intra-pixel crosstalk as a mild convolution with a $3 \times 3$ blurring kernel. We observed that these measures not only helped to improve model immunity to imaging error, but also improved overall model accuracy and robustness to photon noise by reducing overfitting with regularization. The data augmentation was only performed on the input layer rather than all hidden layers, due to computational complexity and the observation that hidden layers were usually more sparse, making crosstalk between neighboring pixels was less of an issue.
    \item Optimizing training parameters: We used a stochastic gradient optimizer for training with a learning rate typically between 0.03 and 0.05, and a momentum between 0.7 and 1. Learning rate decay was applied every 20 epochs with a decaying rate between 0.3 and 0.5. The training parameters were randomly generated within the range for different trials of training, and fine-tuned by using the package Optuna \cite{optuna_2019}.
    
\end{enumerate} 

It is important to note that the quantization of neuron activations was only performed during training on a digital computer, but not during the inference stage on ONN. We observed that even though the models were trained with noise, they still performed better when run with full precision. Therefore, the trained weights and neural activations were loaded with the maximum allowable precision for the hardware (i.e., 7 bits for the OLED display, and 8 bits for the SLM). The training code can be found in this GitHub repository: \url{https://github.com/mcmahon-lab/ONN-QAT-SQL}. 

\section{Workflow for Running Optical Neural Networks for Inference}

\begin{figure}[h!]
\includegraphics [width=0.92\textwidth] {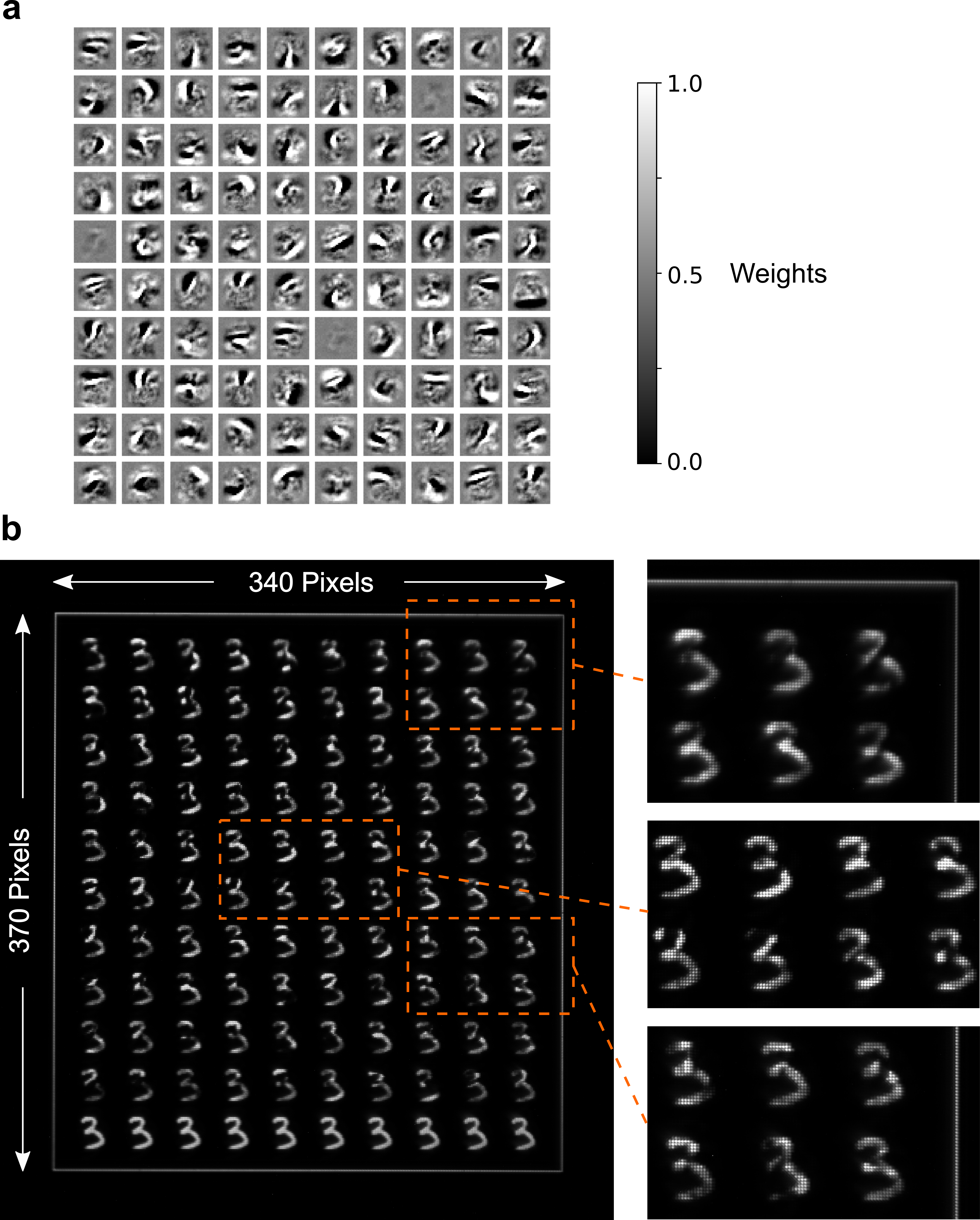}
\caption{\textbf{An Example of the Matrix-vector Multiplication Results of the First Fully-connected Layer of the ONN.} The images show the results of element-wise multiplication between an input vector ($28 \times 28 = 784$ elements) and each row of the matrix of the first fully-connected layer ($100 \times 784$) of our ONN model, as captured by a camera. The last row computed $\sum_{j}x_j$, which was used to offset the output vector for negative elements. Only the output of the first layer is shown. }
\label{digit_mod}
\end{figure}

The trained neural network was executed with optical matrix-vector multiplication in the following steps:
\begin{enumerate}
    \item Starting from the input layer, the matrix-vector multiplication involved in the forward propagation from the current layer to the next layer was computed optically, according to the procedure described in Section \ref{mvm_conversion}. The matrix weights loaded onto the SLM were exactly the same as those in the neural network trained on the digital computer. The number of photons per multiplication in matrix-vector multiplication was controlled by adjusting the number of detector samples to sum. 
    
    \item The bias terms and the nonlinear activation function were applied digitally to the matrix-vector multiplication result, and these parameters followed exactly as those in the trained neural network, without any modification or retraining. The resulting neuron activations were used as the input vector to the matrix-vector multiplication that leads to the next layer (go back to step 1) unless there is none (go to step 3).
    
    \item At the output layer, the prediction was made based on the highest score. 
\end{enumerate} 

For step 1, since the inputs and neural activations were both non-negative due to our choice of ReLU nonlinearity, only the weight matrices needed to be shifted and normalized. The element-wise multiplication of the first layer is visualized in Fig. \ref{digit_mod}b, with the weight matrix displayed on the SLM visualized in Fig. \ref{digit_mod}a for comparison. The $\vec{1} \cdot \vec{x} = \sum_j x_j$ term in Eq. \ref{eq:dot_product_sum} was computed by adding an additional input vector block and setting all the corresponding SLM pixels' transmissivity to the maximum (e.g., Fig. \ref{digit_mod}b, last row in the image. An entire row was used for illustration purposes and redundancy, while in fact only one additional block was needed for the entire layer.) To obtain the answer to the matrix-vector multiplication, element-wise modulated spatial modes in each block were summed up by optical fan-in as described in Section \ref{optical_fan_in}, which is equivalent to summing all the pixels in each block shown in the image taken by the camera in Fig. \ref{digit_mod}. The integrated optical energy was translated to the answer of the dot product according to a calibration curve, which was made by fitting the measured optical energy to the ground truth of the dot products using the first 10 samples of the MNIST test dataset, in a fashion similar to that described in Section \ref{dprod_calibration}. 

For each forward propagation of the neural network, $784 \times 100 + 100 \times 100 + 100 \times 10 = 89,400$ multiplications and 89,400 additions were performed optically. The total digital assistance for each forward propagation include $100+100+10=210$ additions for applying the $\vec{1} \cdot \vec{x}$ term (to shift the dot product between vectors of non-negative elements in order to obtain the dot product between vectors of signed elements),  $100+100+10=210$ additions for applying digital biases, and $100+100=200$ applications of ReLU nonlinear activation functions (involving only simple operations, such as comparing each element to 0 and setting an element to 0 if it is smaller than 0).

\clearpage

\section{Energy Scaling of Optical Neural Networks} \label{scaling}

The total energy consumption, including optical and electrical contributions, has been analyzed in detail for various optical computing/communication systems in the literature \cite{miller2017attojoule, tait2019silicon, hamerly2019large}. Based on the methodology of previous studies, here we discuss the energy consumption of near-future designs of spatially multiplexed optical matrix-vector multipliers. 

\begin{figure}[h!]
\includegraphics [width=0.9\textwidth] {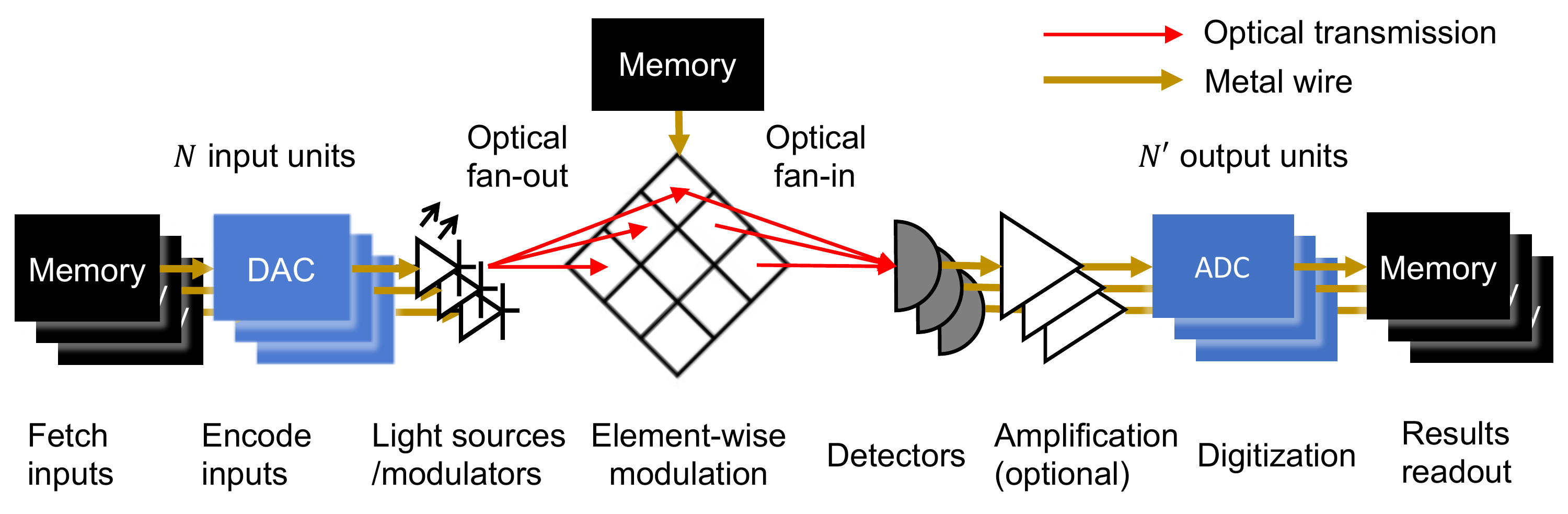}
\caption{\textbf{System Block Diagram of an Integrated Optical Matrix-vector Multiplier.} }
\label{omvm_block_diagram}
\end{figure}

The total energy consumption of a general optical matrix-vector multiplier can be broken down according to its constituent components: light sources, light modulators, detectors for optical signal, detector signal amplifiers, analog-to-digital converters (ADCs), digital-to-analog converters (DACs), and electronic data storage and movement (Fig. \ref{omvm_block_diagram}). The energy cost associated with each part belongs to one of the the two categories of energy scaling, with only one exception that will be pointed out in later discussion: 

\begin{enumerate}
    \item Energy cost that scales with the dimension of input (output) vector size $N$ ($N'$). In this case, the energy consumption per operation scales with $N/(NN')=1/N'$ or $N'/(NN') = 1/N$. Therefore, the energy consumption per operation decreases with the size of matrix-vector multiplication.
    \item Energy cost that scales with the size of the matrix $NN'$. In this case, the energy consumption per operation does not change with $N$ or $N'$. Therefore, such energy costs cannot be amortized with larger matrix-vector multiplications.
\end{enumerate} 

For a near-future blueprint, we are targeting a system performing matrix-vector multiplication of size 4,096$\times$4,096 (i.e., $N=N'=\text{4,096}$) at a \SI{1} {\giga\hertz} update rate with low-precision arithmetic (4-5 bits for both matrix weights and vector activation). The \SI{1} {\giga\hertz} rate can be readily achieved for input data encoding and output results readout with technologies regularly used in telecommunications applications. Electro-optic modulators can achieve up to \SI{\sim 100} {\giga\hertz} in integrated devices compatible with CMOS platforms \cite{wang2018integrated}. Here, we evaluate energy efficiency in terms of energy per multiply-and-accumulate (MAC), which consists of one multiplication and one addition operation as commonly used in deep learning literature \cite{sze2017efficient}. 

With proper implementation, most electronic operations fall into category 1 in terms of energy scaling. For example, the energy consumed by signal amplification, ADCs, memory, and data movement all scales with the output vector size $N'$, and the energy consumed by DACs scales with input vector size $N$. We discuss the energy consumption of each system component in more details below:

\begin{itemize}
    \item Transimpedance amplification is usually required to convert the current signal of photodiodes to voltage; however, if the capacitance of photodiodes is sufficiently small (e.g., \SI{1} {\femto\farad}, which allows the voltage across the capacitor to be \SI{1} {\volt} with \SI{\sim e4}{} electron charges), the voltage can be high enough to switch subsequent ADC circuits without amplification \cite{miller2017attojoule, nozaki2016photonic, tang2008nanometre}. In this case, the energy consumption for each charge transfer is on the order of $CV_{DD}^2 \sim \SI{1} {\femto\joule}$, or $\SI{1} {\femto\joule}/\text{4,096} = \SI{0.24} {\atto\joule\per MAC}$ for $N'=\text{4,096}$. 
    
    \item An ADC is required to convert the voltage signal of each detector to a digital readout for further processing steps, such as applying bias and nonlinear activation functions in neural networks. A high-rate ADC can achieve \SI{\sim 24} {\giga\hertz} conversion rate at \SI{\sim 1} {\pico\joule} per sample for low precision numbers (e.g., 4.5 bits) \cite{ADC2020, xu201623mw}, which means the energy efficiency of the ADC operating at \SI{1} {\giga\hertz} is at least $\SI{1}{\pico \joule}/24=\SI{42} {\femto\joule}$ per sample (since an ADC operating at a higher sampling rate is usually less energy-efficient, due to increased thermal noise). The ADC energy expense per MAC can therefore be reduced to $\SI{42} {\femto\joule}/\text{4,096}=\SI{10} {\atto\joule}$ for $N'=\text{4,096}$. 
    
     \item A DAC is required to convert the digital format of each input vector entry to the current (voltage) controlling the active light source (passive light modulator) that encodes inputs to the system. In existing commercial photonic processors, DAC costs about the same amount of energy per sample as ADC \cite{ramey2020silicon}. In similar electric crossbar applications (5-bit resolution at \SI{100} {\mega\hertz}), each DAC sample has been estimated \cite{cosemans2019towards} to consume as little as \SI{15} {\femto\joule} per sample (\SI{3.7} {\atto\joule\per MAC}).
     
     \item The data transmission in metal wires costs \SI[per-mode=repeated-symbol]{\sim 100}{\atto\joule\per\micro\metre \per bit} on-chip \cite{miller2017attojoule}. With judicious wire planning, it is possible to keep the total energy associated with each MAC on the order of \SI{100}{\atto\joule}, since optical transmission covers quite a bit of distance, which signals would otherwise traverse electronically. 
     
     \item The memory is used to store the detector readouts for digital operations and as inputs to the next layers. The energy cost is usually on the order of \SI{1} {\pico\joule \per bit} for off-chip memory \cite{sze2017efficient}. Therefore, without any optimization, the energy costs associated with data storage is relatively high compared to other costs, at \SI{4} {\pico\joule} per 4-bit number. However, much of this energy is probably consumed for signal transmission during memory access, and the amount of energy to read/write each bit of information in transistors can be reduced to the level of \SI{10} {\femto\joule}, based on the switch energy of transistors \cite{caimi2021scaled}. With tight integration of memory units, the associated energy cost can be hopefully reduced to \SI{100} {\femto\joule} per 4-bit number, which probably has already been achieved based on the total energy efficiency reported in existing digital machine learning accelerators (\SI{10} {\femto\joule \per MAC}-\SI{10} {\pico\joule \per MAC}) \cite{reuthersurvey2020}. Therefore, after amortization, the memory cost can be reduced to the order of $2 \times \SI{100 }{\femto\joule}/\text{4,096} \sim \SI{50 }{\atto\joule \per MAC}$ ($2 \times$ accounts for both input and output).
\end{itemize} 
Overall, with reasonably large input vector sizes (e.g., $N=N'=\text{4,096}$), it is possible to keep energy costs in category 1 on the order of \SI{100} {\atto\joule \per MAC}.

The energy consumption associated with matrix weight modulation scales with the number of entries in the matrix (category 2). For reconfigurable spatial light modulators, changes to the weights are either accomplished all at once with a single investment of energy (e.g., phase-change materials, ferroelectric materials, mechanical phase modulators, etc.), or must be maintained by constant power consumption (e.g., liquid crystal display, LCD). Based on the LCDs equipped on existing commercial mobile devices, it is possible to power millions of pixels with \SI{<1}{\watt} power. The update rate of commercial LCDs is \SIrange[range-phrase=-]{60}{240}{}\SI{}{\hertz}. For machine-learning inference applications, where the weights can stay stationary most of the time, the energy cost per multiplication can be amortized by increasing the update rates of system input and output devices. For example, with a 16-million-pixel LCD consuming \SI{1}{\watt} with stationary weights, if other parts of the system update at \SI{1}{\giga\hertz}, the energy consumption for each multiplication would be $\SI{1}{\watt}/(\SI{e9}{MACs \per second})/(\text{4,096} \times \text{4,096}) = \SI{62.5}{\atto\joule \per MAC}$.

The energy scaling of the light source and detector is in general more complicated, depending on whether the system is working in a shot noise-limited regime (category 1 for fixed output precision), or a thermal-noise-limited regime (in-between categories 1 and 2 with weaker energy benefits for larger $N$ or $N'$, see Section \ref{optical_fan_in}). Here, we argue that no matter which case is true, the optical energy consumption is insignificant compared to the electronic energy costs analyzed above. As discussed in Section \ref{optical_fan_in}, at extremely low photon fluxes (\textless1 photon per multiplication), the thermal noise becomes a more prominent factor in determining detection SNR. Even though the detector gain can be increased to improve SNR, such measures would not improve the overall energy efficiency, since it is more economical to have stronger signal with more photons, than amplifying photoelectrons. Therefore, in a realistic device, a few photons (\SI{\sim e-18} {\joule}) should be budgeted for each multiplication at the detector in order to ensure sufficiently high SNR. With optical fan-out, such a low photon budget for each spatial mode can be achieved by distributing photons generated by a single, intense light source to a large number of $N$ spatial modes. The use of a single light source substantially reduces the number of light sources that needs to be integrated and the energy cost spent on distributing data among the light sources through electronic data transmission \cite{miller2017attojoule}. Therefore, a 10\% ``wall-plug" efficiency can be realistically expected from an active coherent light source array (e.g., VCSEL), or a passive modulator array (e.g., thin-film lithium niobate modulators) with a single external laser source. Incidentally, the additional energy for operating the passive modulator array can be reduced to \SI{<10} {\femto\joule} for 4-bit numbers running at \SI{>1} {\giga\hertz} \cite{wang2018integrated}. The energy per MAC is then $\SI{<10}{\femto\joule}/\text{4,096} = \SI{2.5}{\atto\joule \per MAC}$, which is insignificant compared to other energy costs. Overall, it is reasonable to expect the lumped quantum efficiency, including the light source and detector, to achieve 1\%, and thus the electrical energy cost per MAC would be $\SI{e-18}{\joule}/1\% = \SI{100}{\atto\joule}$.

In summary, based on currently available technologies, it is realistic to build an optical matrix-vector multiplier performing matrix-vector multiplication of size $\text{4,096} \times \text{4,096}$ at an update rate of \SI{1} {\giga\hertz} reaching a speed of \SI{\sim 17e15} {MACs \per \second} (or 17 POPS). The energy consumption for each MAC can be on the order of several hundred \SI{} {\atto \joule} (excluding memory), or \SI{1} {\femto\joule} (including memory), which leads to a total power of \SI{\sim 40} {\watt} (including memory). The energy efficiency will be 2 to 3 orders of magnitude better than the state-of-the-art digital electronic machine learning accelerators (e.g., \SI{10} {\femto\joule \per MAC}-\SI{10} {\pico\joule \per MAC}) \cite{reuthersurvey2020, irds_2020} or photonic chips \cite{ramey2020silicon}, and better than the blueprint for analog electronic computing \cite{cosemans2019towards}. In the case where the average power is limited by heat dissipation rate, energy efficiency translates to computational speed, which means the optical processor can achieve computation speeds $10 \times$ to $100 \times$ higher than digital processors with the same average power. 

It is worthwhile to reiterate that the optical energy advantage stems from low-loss optical communication over a distance of \SIrange[range-phrase=-]{10}{100}{} \SI{}{\micro\metre} for on-chip platforms \cite{miller2017attojoule}. This enables spatial multiplexing in a region much larger than what is possible with pure electronic circuits, which would either reach the limit of heat dissipation by scaling down element sizes or result in most of the energy being spent on communication in a large area. As a result, optical platforms allow the matrix size to scale up more easily, to the level of $10^3 \times 10^3$ or even $10^4 \times 10^4$, limited only by the number of light sources, modulators, and detectors that can be integrated. In addition, optical operations such as fan-in and fan-out feature extremely high levels of branching ($10^3$-$10^4$) that are not practical for electronics, which help to optimize the data flow and to save a significant portion of energy that would otherwise be spent on storing or relaying intermediate results (for more details, see Section \ref{optical_fan_in}). Overall, optical operations not only allow the energy cost per MAC to scale down with the size of optical matrix-vector multiplication, but also has a short-term potential to reach a sufficiently large matrix size in order to achieve competitive energy efficiency. 

\clearpage
\bibliographystyle{pnas_modified.bst}
\bibliography{references}